\documentclass[12pt]{article}
\usepackage{amssymb,amsmath,amsfonts,eurosym,geometry,ulem,graphicx,caption,color,setspace,sectsty,comment,footmisc,caption,natbib,pdflscape,array,hyperref}

\usepackage{float}        
\usepackage{threeparttable} 
\usepackage{multirow}    
\usepackage{subcaption}
\usepackage{authblk} 
\usepackage{graphicx}
\usepackage{caption}
\usepackage{enumitem}

\usepackage{alphalph}

\hypersetup{
	colorlinks=true,    
	linkcolor=blue,       
	citecolor=blue,       
	filecolor=blue,       
	urlcolor=blue        
}

\newcolumntype{L}[1]{>{\raggedright\let\newline\\arraybackslash\hspace{0pt}}m{#1}}
\newcolumntype{C}[1]{>{\centering\let\newline\\arraybackslash\hspace{0pt}}m{#1}}
\newcolumntype{R}[1]{>{\raggedleft\let\newline\\arraybackslash\hspace{0pt}}m{#1}}

\begin{document}

\begin{titlepage}
\title{An Empirical Comparison of Weak-IV-Robust Procedures in Just-Identified Models}
\author{
  Wenze Li\thanks{Division of Economics, School of Social Sciences, Nanyang Technological University. Email: \texttt{wenze001@e.ntu.edu.sg}.\\
  
  I am sincerely grateful to Wenjie Wang (Nanyang Technological University) and Liyu Dou (Singapore Management University) for their generous guidance and insightful comments, which have substantially strengthened the arguments and analysis presented in this paper.}
}

\date{\today}
\maketitle
\begin{abstract}
\noindent
Instrumental variable (IV) regression is recognized as one of the five core methods for causal inference, as identified by \citet{Angrist-Pischke(2008)}. This paper compares two leading approaches to inference under weak identification for just-identified IV models: the classical Anderson--Rubin ($AR$) procedure and the recently popular \( tF \) method proposed by \citet{lee2022valid}. Using replication data from the \textit{American Economic Review} (AER) and Monte Carlo simulation experiments, we evaluate the two procedures in terms of statistical significance testing and confidence interval (CI) length. Empirically, we find that the $AR$ procedure typically offers higher power and yields shorter CIs than the \( tF \) method. Nonetheless, as noted by \citet{lee2022valid}, \( tF \) has a theoretical advantage in terms of expected CI length. Our findings suggest that the two procedures may be viewed as complementary tools in empirical applications involving potentially weak instruments.


\bigskip
\end{abstract}

\end{titlepage}

\section{Introduction}

The instrumental variable (IV) regression is one of the five most commonly used causal inference methods identified by \cite{Angrist-Pischke(2008)}.
However, weak instruments remain persistent concerns in IV regressions across various fields. 
For instance, surveys by \cite{Andrews-Stock-Sun(2019)}, \cite{young2022}, and \cite{lee2022valid} find that a considerable number of IV specifications in the \textit{American Economic Review} (AER) report first-stage $F$-statistics below 10. 

Consider the commonly used linear IV model, with outcome $Y$, endogenous regressor $X$, and instrument $Z$:
\begin{align}
	& \text{Second stage:} \ Y = X\beta + u \label{eq:second_stage} \\
	& \text{First stage:}  \quad X = Z\pi + v \label{eq:first_stage},
\end{align}
\noindent
where \( Cov(u, Z) = 0 \) and \( Cov(Z, X) \neq 0 \).
The problem of conducting hypothesis tests and constructing confidence sets for the structural parameter $\beta$ with valid significance and confidence levels has been extensively studied for decades. In this context, the Anderson–Rubin ($AR$) test remains a well-established and foundational method (e.g., see \cite{anderson1949estimation}, \cite{Dufour(1997)}, and \cite{Andrews-Stock-Sun(2019)}, among others). In particular, a variety of weak-identification-robust procedures have been built upon $AR$ (e.g., see 
\cite{Stock-Wright(2000)}, \cite{Kleibergen(2002)}, \cite{Moreira(2003)}, \cite{Andrews(2016)}, and \cite{Andrews-Mikusheva(2016)}).
The $AR$ test is uniformly valid under the weak instrument asymptotics of \cite{Staiger-Stock(1997)} and maintains correct size regardless of the strength of the first-stage regression. Furthermore, \cite{Moreira(2009)} shows that $AR$ is the uniformly most powerful unbiased test in the just-identified case. 

Recently, \citet{lee2022valid} propose a new 
weak-IV-robust method for just-identified models, termed the $tF$ procedure, which preserves the interpretability of the conventional 2SLS t-ratio while addressing its invalidity under weak identification. Building on the framework of \citet{stock2002testing}, the $tF$ procedure introduces a smooth critical value (CV) function indexed by the first-stage $F$-statistic, thereby ensuring uniform size control across different levels of instrument strength. Furthermore, the authors argue that $tF$ improves upon the $AR$ procedure in certain respects—most notably by yielding confidence intervals (CIs) with shorter expected lengths than those of $AR$, whenever both are bounded.

However, empirical evidence regarding the relative performance of the two weak-IV-robust procedures remains sparse. In this paper, we conduct a performance comparison regarding the test of statistical significance and CI length between the two procedures by using both empirical dataset and simulation experiments. We find that the $AR$ procedure tends to outperform the $tF$ procedure in several key aspects.

\section{Performance Comparison Between $AR$ and \( tF \)}
\subsection{Dataset for the Empirical Comparison}

Our empirical analysis is based on the same AER dataset as that studied in  \citet{lee2022valid}, which includes all AER articles published between 2013 and 2019, excluding comments, replies, and AEA Papers and Proceedings. This yields a total of 757 articles, among which 123 contain IV regressions. Of these, 61 studies employ single-instrument (just-identified) regressions, contributing a total of 1,311 specifications of IV regressions to the dataset. Among them, 458 specifications across 39 studies have matching sample sizes in the first- and second-stage regressions. Another 19 specifications are dropped due to missing the values of $F$-statistics, resulting in a working dataset of 439 specifications from 39 studies, as used by \citet{lee2022valid}.

However, we find that this dataset still contains several cases with multiple instruments or nonlinear model structures, which may violate the assumptions underlying the $AR$ and \( tF \) procedures. After removing these cases, 343 specifications from 36 studies remain. 
Among the remaining specifications, we are able to compute the $AR$ test for 151 specifications from 17 studies, primarily due to limitations related to data confidentiality (i.e., we exclude the studies that are based on private data). 
Additional details of the 17 studies are provided in the Online Appendix. 

\subsection{Statistical Significance}

\begin{figure}[H]
	\centering
	\includegraphics[width=0.8\textwidth]{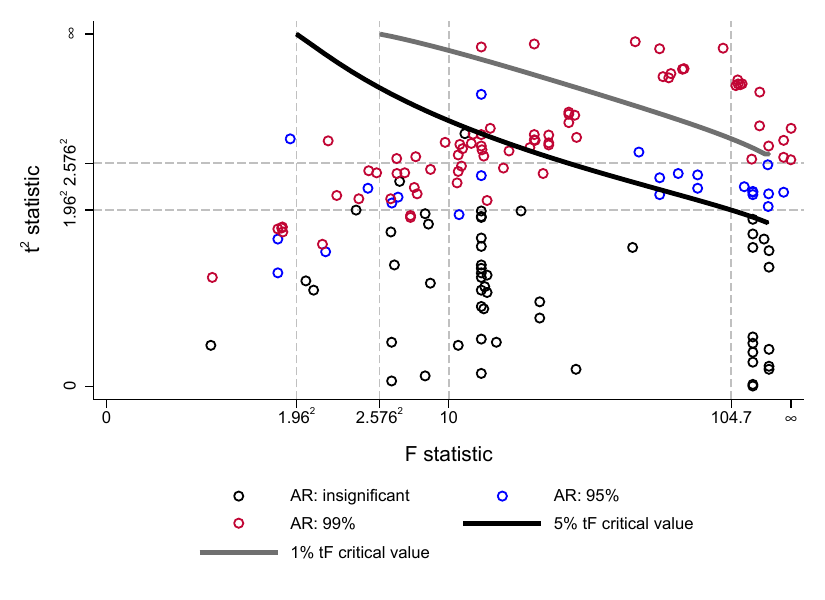}
	\caption{Statistical Significance of $t$-ratio, $AR$, and $tF$}
    \caption*{\fontsize{8pt}{10pt}\selectfont \textit{Notes:} This figure is based on 151 specifications. The vertical axis plots \( \frac{t^2 / 1.96^2}{1 + t^2 / 1.96^2} \), while the horizontal axis plots \( \frac{F / 10}{1 + F / 10} \). For the $AR$ test, black circles denote insignificance, blue circles indicate significance at the 5\% level only, and red circles represent significance at the 1\% level. The solid black line represents the 5\% CV for the \( tF \) test, and the solid gray line represents the 1\% level.}
	\label{fig:1}
\end{figure}

We first test the null hypothesis $H_0: \beta = 0$ for all the 151 specifications. To compare the null rejection behavior of the $AR$ and \( tF \) tests using the AER replication sample, we present Figure \ref{fig:1}. 
Let $t$ denote the conventional $t$-ratio statistic. 
For each specification in our dataset, the vertical axis plots the value of its standardized \( t^2 \) statistic, defined as \( \frac{t^2 / 1.96^2}{1 + t^2 / 1.96^2} \), 
while the horizontal axis shows the value of the corresponding standardized first-stage \( F \)-statistic, calculated as \( \frac{F / 10}{1 + F / 10} \), to facilitate full visualization of both statistics. Additionally, we plot the conventional rejection thresholds for the $t$-ratio statistics in dotted lines (that is, \( t^2 > 1.96^2 \) for the 5\% significance level and \( t^2 > 2.576^2 \) for the 1\% level, respectively). 

In Figure \ref{fig:1}, black circles indicate insignificance, blue circles denote significance at the 5\% level only, and red circles represent significance at the 1\% level, respectively, for the $AR$ procedure.  
Then, we plot CV curves for the \( tF \) procedure in Figure \ref{fig:1}, where the solid black line represents the CV curve for the 5\% level and the solid gray line represents that for the 1\% level, respectively.  
The $tF$ CV depends on the values of both $t^2$ and $F$ by construction. 

We observe that among the specifications deemed insignificant under the \( tF \) procedure, 47.52\% are nonetheless found to be significant under the $AR$ test—corresponding to the proportion of blue and red circles located below the 5\% \( tF \) CV curve (solid black line). Furthermore, among those specifications that are significant at the 5\% level under the \( tF \) procedure but not at the 1\% level, 50\% are in fact significant at the 1\% level under $AR$—represented by the proportion of red circles within the region bounded between the 5\% and 1\% \( tF \) CV curves (i.e., the area between the solid black and gray lines). Importantly, there are no cases in which a specification is significant under the \( tF \) procedure at a level more stringent than under $AR$, reflecting the relatively conservative nature of the \( tF \) test for testing statistical significance in the current dataset. These findings suggest that the $AR$ procedure may have a power advantage over the $tF$ procedure in a variety of empirically relevant cases.

\subsection{Power Simulations}

To further examine the power properties of the two procedures, we conduct simulation experiments, systematically comparing the performance of the $AR$ and \( tF \) procedures under a range of scenarios. Specifically, we plot power curves under a wide range of alternatives \( \beta - \beta_0 \), where \( \beta \) denotes the true parameter value and \( \beta_0 \) the null hypothesis. Rejection probabilities are evaluated as a function of the deviation \( \beta - \beta_0 \) across various combinations of the key nuisance parameters \( \rho \) and \( f_0 \). 
In particular, instrument strength is governed by \( f_0 \) via the relationship \( \mathbb{E}[F] = f_0^2 + 1 \), while the degree of endogeneity is captured by \( \rho \), defined as the correlation between the first-stage and structural regression residuals. 

Further details of the simulations and the power curves are presented in the Online Appendix. We highlight several main findings below. 
First, the power results indicate that the $AR$ test demonstrates higher power than the \( tF \) procedure in most cases, particularly when the instrument strength is low (e.g., when $f_0$ is equal to 1, 2, or 4).
Second, across all simulations, the power of the $AR$ test appears to (weakly) dominate that of the $tF$ test when the endogeneity level is relatively low (e.g., when $\rho$ is equal to -0.3, -0.1, 0.1 or 0.3). Third, the $AR$ test has a substantial power advantage for testing positive alternatives in the cases with a negative $\rho$, while having an advantage for testing negative alternatives in the cases with a positive $\rho$.

\subsection{Confidence Interval Lengths}

In this section, we compare CI lengths of  the $AR$ and $tF$ procedures using the AER sample. 
 
\begin{figure}[H]
    \centering

    \begin{subfigure}[t]{0.7\textwidth}
        \centering
        \includegraphics[width=\textwidth]{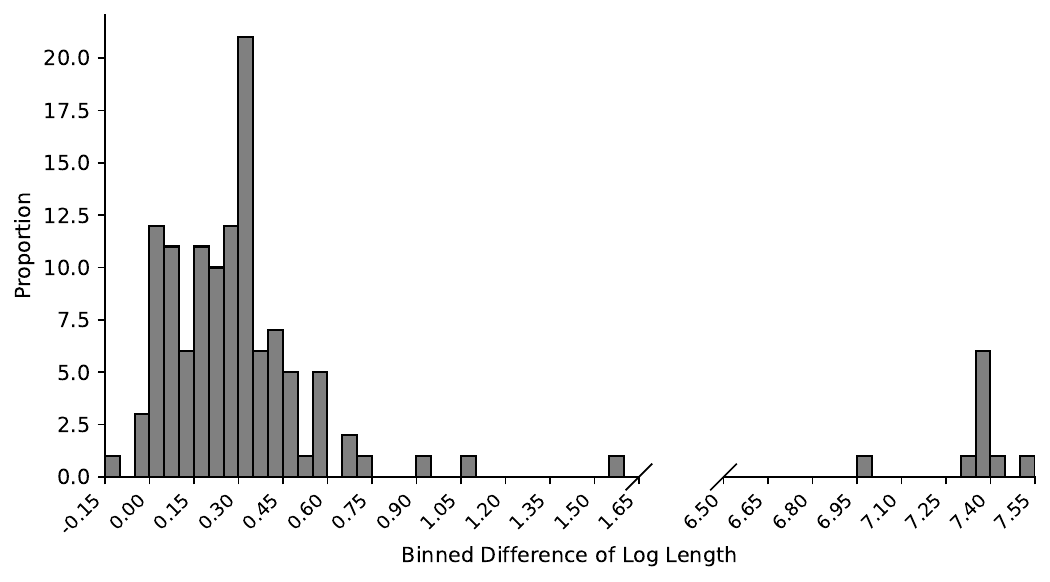}
        \caption{\fontsize{8.5pt}{10pt} 5\% level}
    \end{subfigure}
    \hfill
    \begin{subfigure}[t]{0.7\textwidth}
        \centering
        \includegraphics[width=\textwidth]{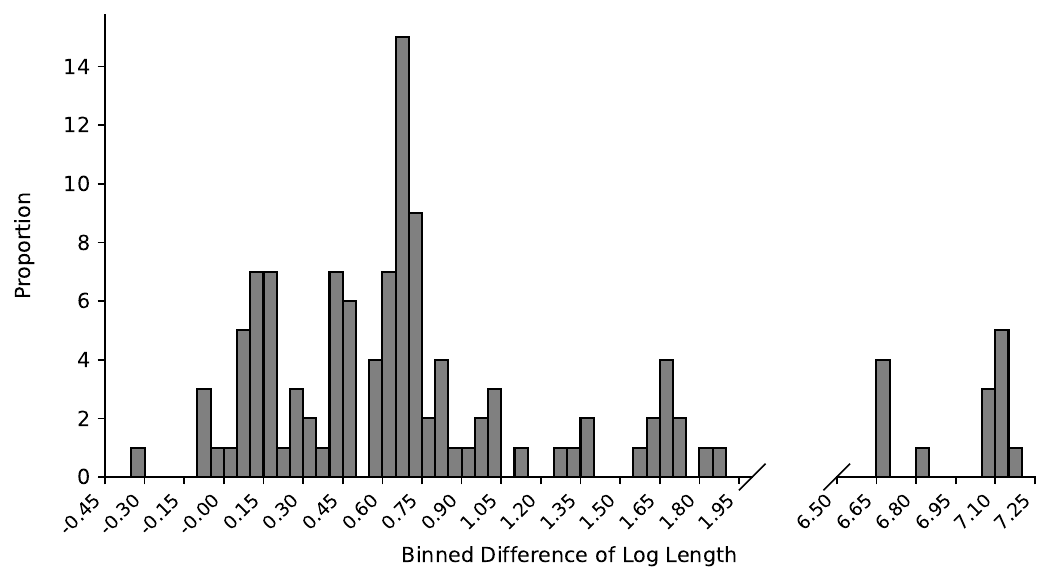}
        \caption{\fontsize{8.5pt}{10pt} 1\% level}
    \end{subfigure}

    \caption{Distribution of $\ln\left( \frac{\text{length}_{tF}}{\text{length}_{AR}} \right)$}
    \caption*{\fontsize{8pt}{10pt}\selectfont \textit{Notes:} Part (a) is based on 127 specifications at the 5\% level, and part (b) on 123 specifications at the 1\% level. Each bar represents the frequency of observations within a 0.05-width bin of the log difference in CI length.}
    \label{fig:2}
\end{figure}

For each specification, we compute the difference in log lengths between the two procedures:
\[
\ln\left( \frac{\text{length}_{tF}}{\text{length}_{AR}} \right).
\]
The distribution of this measure is reported in Figure \ref{fig:2}. 
Among the 151 specifications, one is excluded because \( t_{\text{AR}}^2 = 0 \), which precludes the computation of \( \hat{\rho} \) (needed for the heatmaps in Figure \ref{fig:3}),
and another is removed due to an invalid estimated correlation (i.e., \( |\hat{\rho}| > 1 \)). After excluding cases with missing CI length information, 127 specifications remain at the 5\% significance level and 123 at the 1\% level. 

We find that the $AR$ CI is shorter than that of the $tF$ procedure in 96.85\% of specifications at the 5\% level and 95.93\% at the 1\% level. Among cases in which the $AR$ interval is longer, the average log difference is \(-5.64\) log points at the 5\% level and \(-10.84\) log points at the 1\% level. Conversely, when the $AR$ interval is shorter, the average log difference is 86.41 log points at the 5\% level and 142.39 log points at the 1\% level. Furthermore, in 48.03\% of cases at the 5\% level and 76.42\% at the 1\% level, the $tF$ interval exceeds the $AR$ interval by more than 30 log points. For context, the standard 95\% CI is longer than the 90\% interval by approximately \( \ln\left( \frac{1.96}{1.645} \right) \approx 0.18 \), and the 99\% interval exceeds the 95\% interval by approximately \( \ln\left( \frac{2.58}{1.96} \right) \approx 0.27 \), highlighting the substantive magnitude of these differences. Overall, the $AR$ procedure yields shorter CIs than the $tF$ procedure in the majority of cases. Moreover, in the current sample, when the $AR$ method outperforms, it does so substantially; when it underperforms, the extent of loss is typically modest.

Figure \ref{fig:3} presents a heatmap of the log difference in CI lengths, plotted against two key diagnostic measures: the first-stage \( F \)-statistic (vertical axis) and the absolute value of the estimated residual correlation, \( |\hat{\rho}| \) (horizontal axis). To facilitate visualization, the vertical axis applies the transformation \( \frac{F/10}{1 + F/10} \), which maps the \( F \)-statistic onto the unit interval. Representative values of \( F = 104.67 \), \( 10 \), \( 2.576^2 \), and \( 1.96^2 \) correspond to transformed values of the vertical axis \( y = 0.91 \), \( 0.5 \), \( 0.4 \), and \( 0.28 \), respectively. Each point on the heatmap reflects an interpolated average of the log difference computed over neighboring observations. Red regions indicate positive values of \( \ln\left( \frac{\text{length}_{tF}}{\text{length}_{AR}} \right) \), with deeper shades signifying larger differences in favor of the $AR$ procedure. Conversely, blue regions denote negative values, where the \( tF \) procedure yields shorter intervals. Overall, the figure demonstrates that the $AR$ procedure typically delivers substantially shorter CIs than the \( tF \) procedure, particularly in areas with low first-stage \( F \)-statistics—that is, when the IV is weak.

\begin{figure}[H]
    \centering
    \begin{subfigure}[t]{0.76\textwidth}
        \centering
        \includegraphics[width=0.76\textwidth]{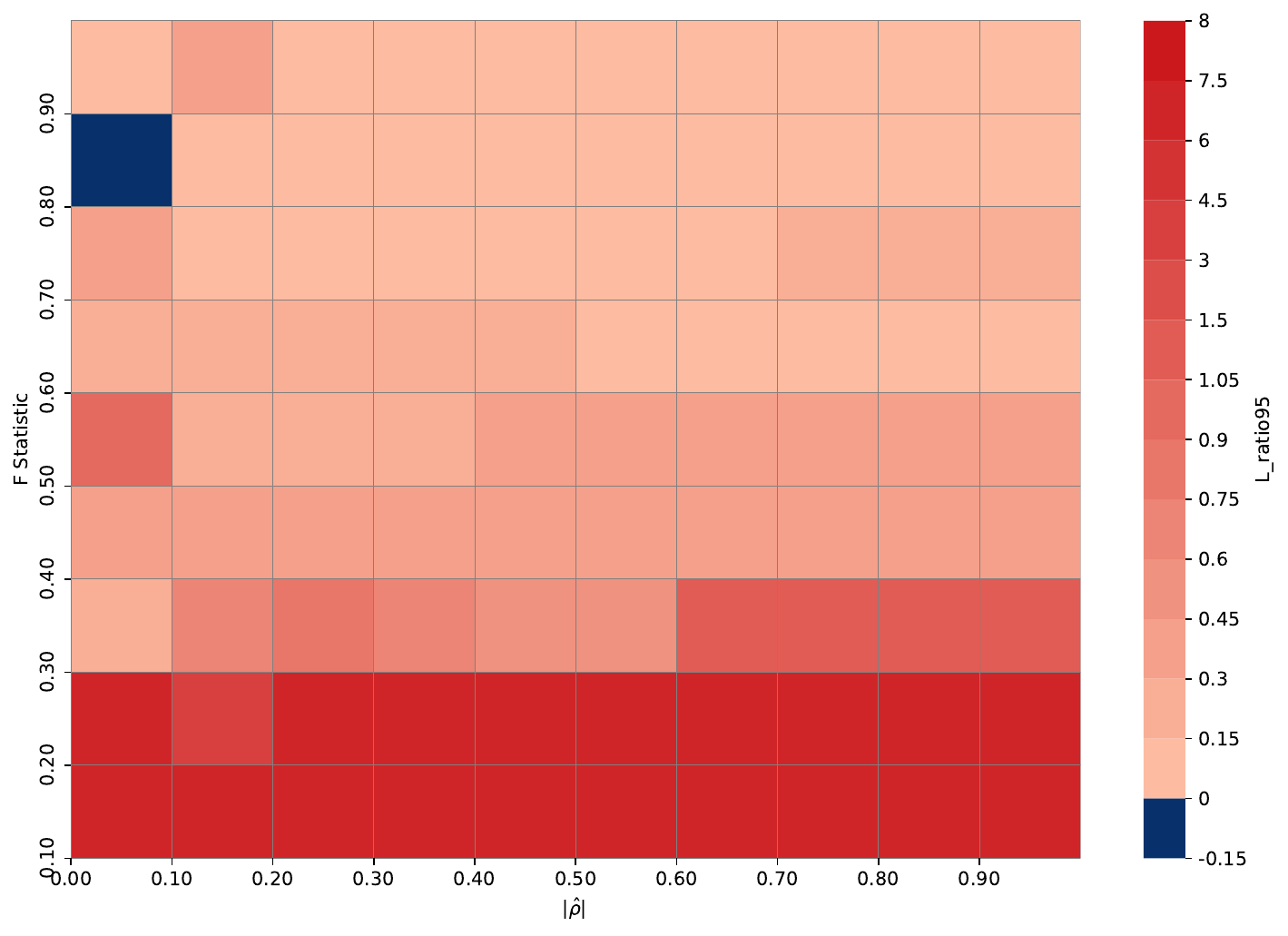}
        \caption{5\% level}
    \end{subfigure}
    \caption{Heatmap of $\ln\left( \frac{\text{length}_{tF}}{\text{length}_{AR}} \right)$}
    \label{fig:3}
\end{figure}

\begin{figure}[H]
    \ContinuedFloat
    \centering
    \begin{subfigure}[t]{0.76\textwidth}
        \centering
        \includegraphics[width=0.76\textwidth]{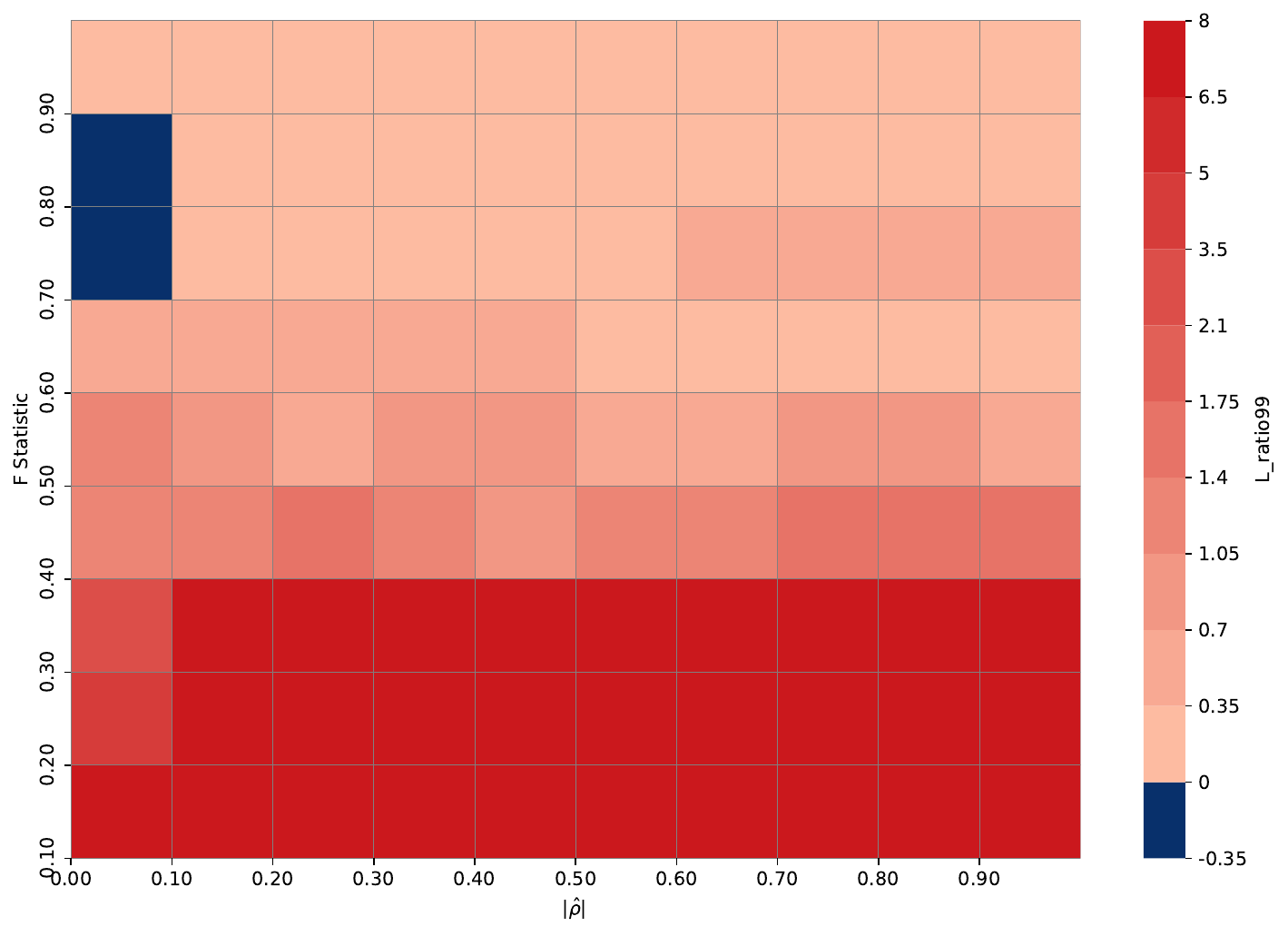}
        \caption{1\% level}
    \end{subfigure}
    \caption{Heatmap of $\ln\left( \frac{\text{length}_{tF}}{\text{length}_{AR}} \right)$ (continued)}
   \caption*{\fontsize{8pt}{10pt}\selectfont \textit{Notes:} The vertical axis applies the transformation \( \frac{F/10}{1 + F/10} \) to map the first-stage \( F \)-statistic onto the unit interval. Under this scaling, \( F \) values of 104.67, 10, \( 2.576^2 \), and \( 1.96^2 \) correspond to vertical positions of approximately 0.91, 0.5, 0.4, and 0.28, respectively. In the heatmap, red regions indicate positive values of \( \ln\left( \frac{\text{length}_{tF}}{\text{length}_{AR}} \right) \), with deeper shades representing larger differences, favoring the $AR$ procedure. Conversely, blue regions denote negative values, indicating instances where the \( tF \) procedure produces shorter CIs.}
\end{figure}

\section{Conclusion}

In this paper, we conduct a performance comparison regarding the test of statistical significance and CI length between the $AR$ and $tF$ procedures by using both AER dataset and simulation experiments. We find that the $AR$ procedure tends to outperform the $tF$ procedure in several key aspects. 
This is also in line with the results in \cite{keane2023instrument, keane2024practical}, who recommend using the $AR$ procedure rather than $t$-ratio based procedures.  
However, we note that as pointed out by \cite{lee2022valid}, $tF$ has the theoretical advantage over $AR$ in terms of expected CI length. Therefore, we believe that the two procedures are complementary to each other. 
There are several potential directions for future research. 
First, \cite{lee2023you} recently proposed a refined $tF$ procedure (the $VtF$ procedure). It is important to empirically investigate the performance of the $VtF$ procedure compared with $AR$. Second, there is a growing literature on weak-identification-robust inference under many weak instruments and non-homoskedastic errors.\footnote{E.g., see \cite{crudu2021}, \cite{MS22}, \cite{matsushita2024jackknife}, \cite{LWZ(2023)}, \cite{DKM24},  \cite{boot-ligtenberg(2023)}, \cite{N23}, and \cite{lim2024dimension}, among others.}  
As pointed out by \cite{yap2023valid}, the asymptotic framework of many weak instruments is closely related to that of just-identified models. It may be, therefore, interesting to extend the empirical investigation to the applications with many weak instruments. Third, the $AR$ and $tF$ procedures studied in the paper are based on asymptotic CVs, which may not have satisfactory finite sample performance under non-homoskedastic errors. On the other hand, it is found that when implemented appropriately, bootstrap approaches may substantially improve the inference accuracy for IV models, including the cases where IVs may be rather weak.\footnote{E.g., see 
\cite{Moreira-Porter-Suarez(2009)}, 
\cite{Davidson-Mackinnon(2008), Davidson-Mackinnon(2010), Davidson-Mackinnon(2014b)}, 
\cite{wang2015bootstrap},
\cite{Wang-Kaffo(2016)}, \cite{Kaffo-Wang(2017)}, \cite{Wang-Doko(2018)}, \cite{Finlay-Magnusson(2019)}, \cite{Roodman-Nielsen-MacKinnon-Webb(2019)}, \cite{mackinnon2023fast},
and \cite{Wang-Zhang2024}.} It may be interesting to consider the bootstrap version of the $tF$ or $VtF$ procedure for performance improvement.

\newpage
\label{Bibliography}
\setstretch{1}
\bibliographystyle{apalike}
\bibliography{reference}

\newpage
\section*{Online Appendix}
\pdfbookmark[1]{Online Appendix}{appendix}

\appendix

\subsection*{A \quad Monte Carlo Simulations}
\pdfbookmark[2]{Monte Carlo Simulations}{simulations}

In this section, we conduct Monte Carlo simulations using the same data generating process as that in the simulations of \cite{lee2022valid}, each based on a sample size of 1{,}000. The number of Monte Carlo replications is equal to 250,000 across the simulations. 
Throughout the experiments, the true value of the structural parameter is fixed at \( \beta = 1 \). We consider a range of instrument strengths, indexed by \( f_0 \in \{1, 2, 4, 6, 8, 10\} \), and values of the endogeneity parameter \( \rho \) varying from \(-0.9\) to \(0.9\) in increments of 0.2. For each pair \( (f_0, \rho) \), we evaluate across a grid of deviations \( \beta - \beta_0 \), ranging from \(-1.6\) to \(1.5\) in steps of 0.1. 
Specifically, for each combination of \( \beta_0 \), \( f_0 \), and \( \rho \), we compute the rejection probabilities and compare the empirical power performance of the $AR$ procedure to that of the \( tF \) test. The results, presented in Figure \ref{fig:A1}, indicate that the $AR$ test demonstrates higher power than the \( tF \) procedure in most cases, particularly when the instrument strength is low (e.g., when $f_0$ is equal to 1, 2, or 4). Additionally, across all simulations, the power of the $AR$ test appears to (weakly) dominate that of the $tF$ test when the endogeneity level is relatively low (e.g., when $\rho$ is equal to -0.3, -0.1, 0.1 or 0.3). Furthermore, we note that the $AR$ test has a substantial power advantage for testing positive alternatives in the cases with a negative $\rho$, while having an advantage for testing negative alternatives in the cases with a positive $\rho$. 

\begin{figure}[H]
	\centering
	
	\begin{subfigure}[t]{0.375\textwidth}
		\includegraphics[width=\linewidth]{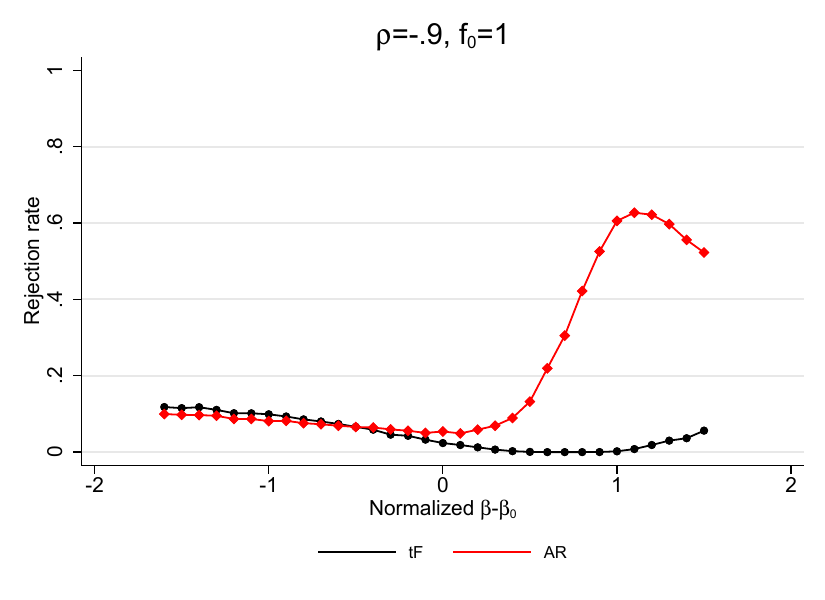}
	\end{subfigure}
	\begin{subfigure}[t]{0.375\textwidth}
		\includegraphics[width=\linewidth]{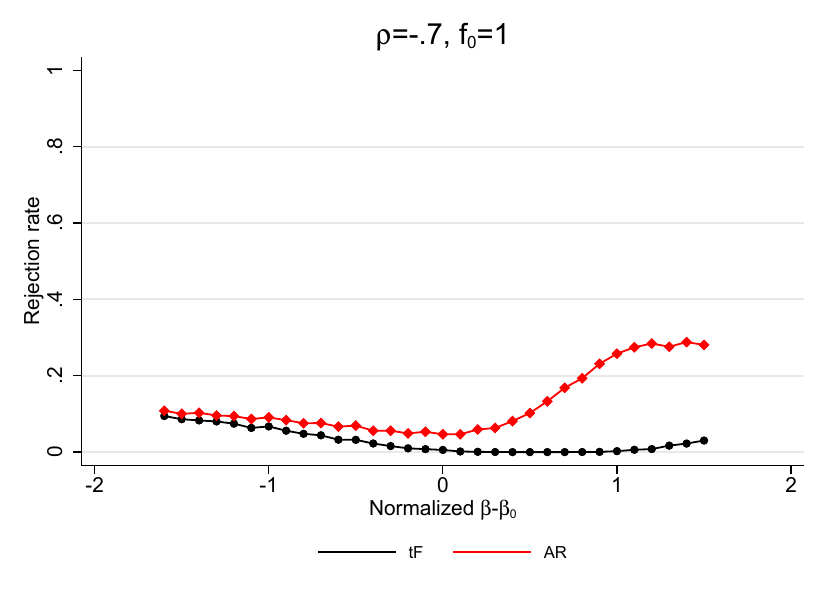}
	\end{subfigure}

	\begin{subfigure}[t]{0.375\textwidth}
		\includegraphics[width=\linewidth]{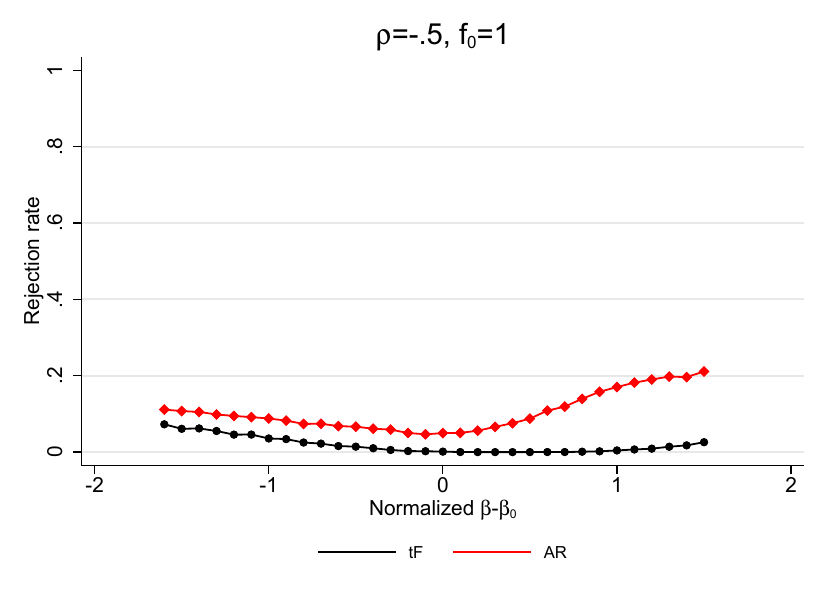}
	\end{subfigure}
	\begin{subfigure}[t]{0.375\textwidth}
		\includegraphics[width=\linewidth]{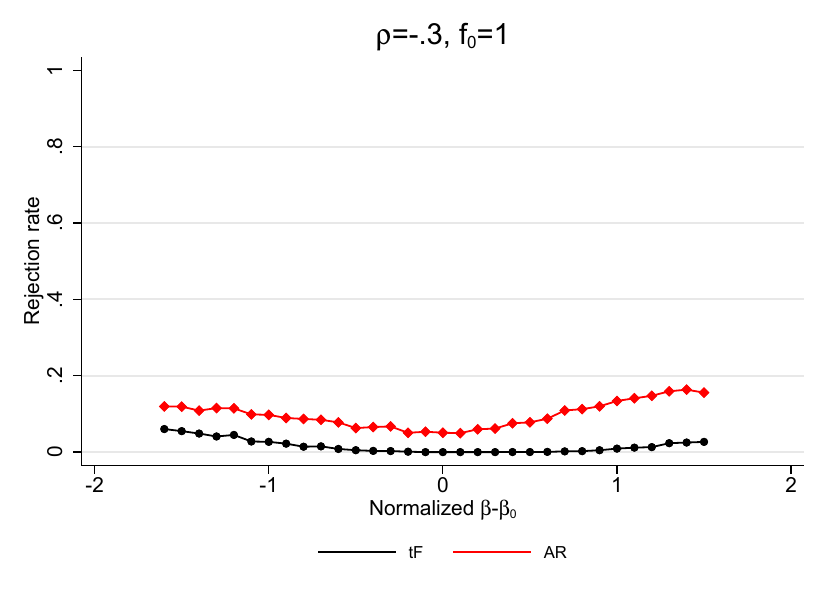}
	\end{subfigure}

	\begin{subfigure}[t]{0.375\textwidth}
		\includegraphics[width=\linewidth]{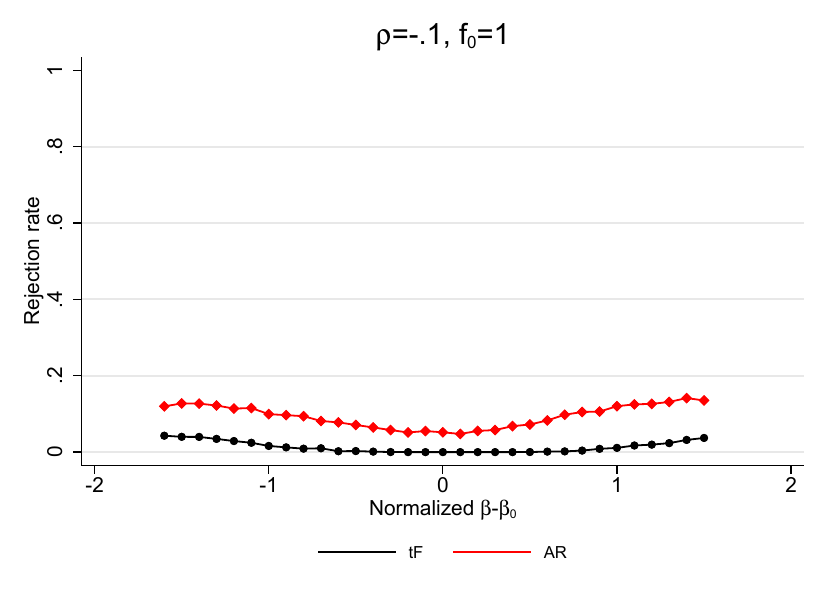}
	\end{subfigure}
    \begin{subfigure}[t]{0.375\textwidth}
		\includegraphics[width=\linewidth]{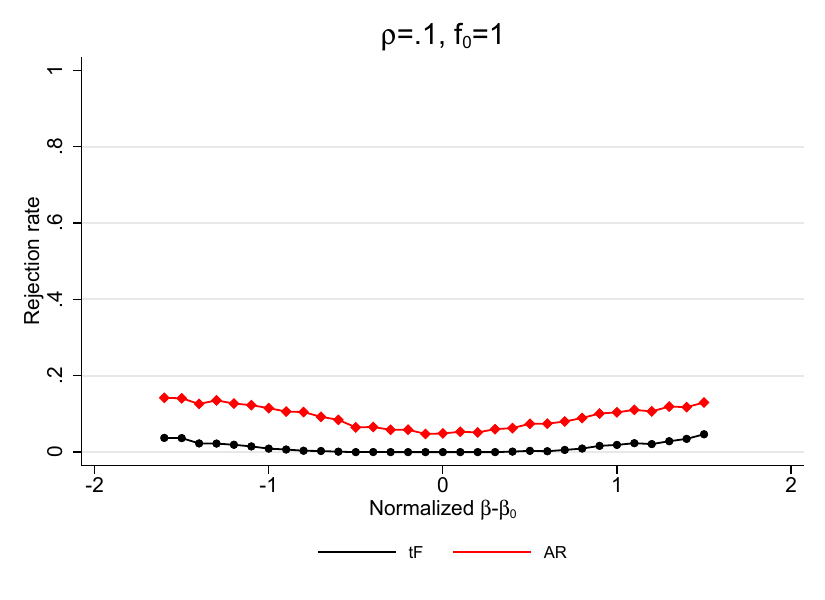}
	\end{subfigure}
    
    \begin{subfigure}[t]{0.375\textwidth}
		\includegraphics[width=\linewidth]{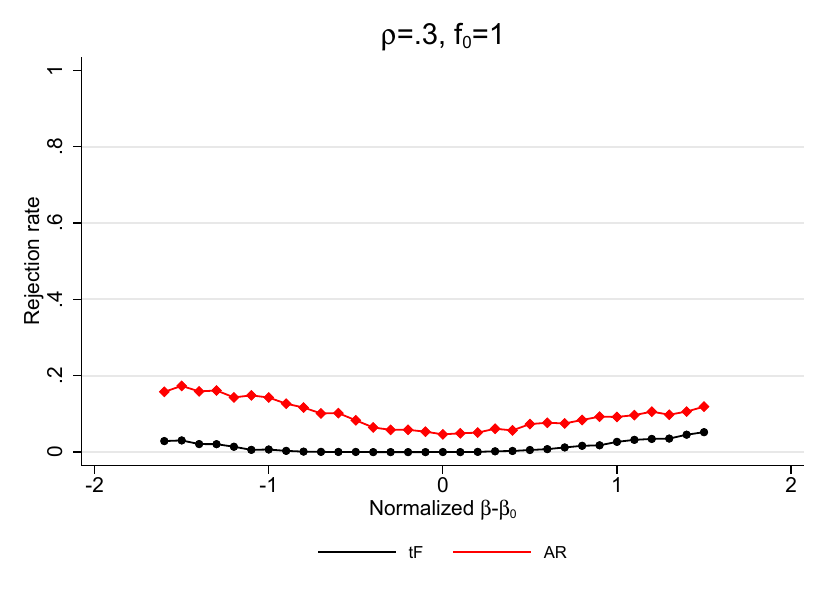}
	\end{subfigure}
    \begin{subfigure}[t]{0.375\textwidth}
		\includegraphics[width=\linewidth]{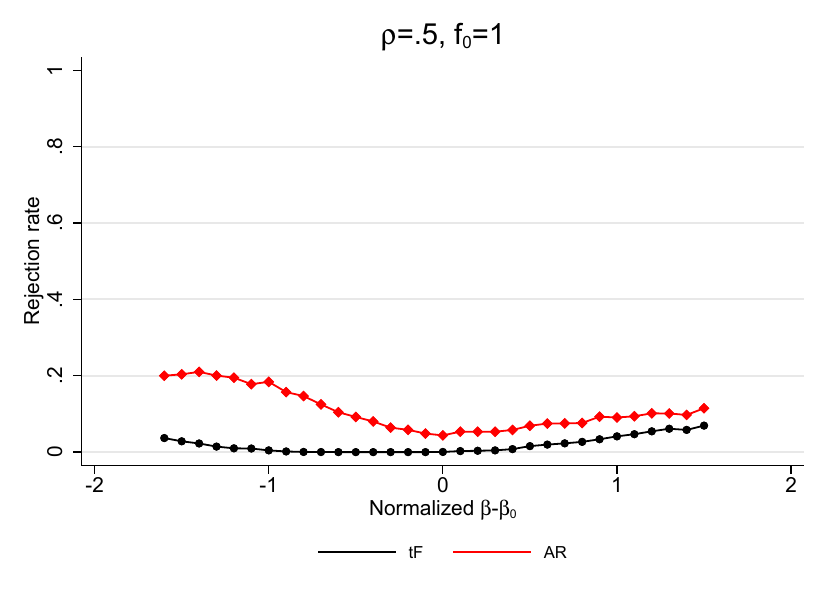}
	\end{subfigure}

    \begin{subfigure}[t]{0.375\textwidth}
		\includegraphics[width=\linewidth]{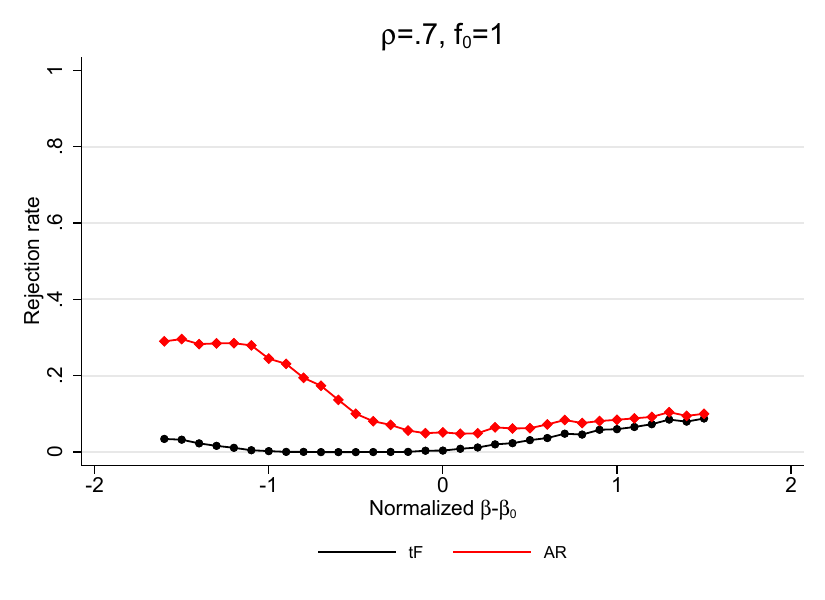}
	\end{subfigure}
    \begin{subfigure}[t]{0.375\textwidth}
		\includegraphics[width=\linewidth]{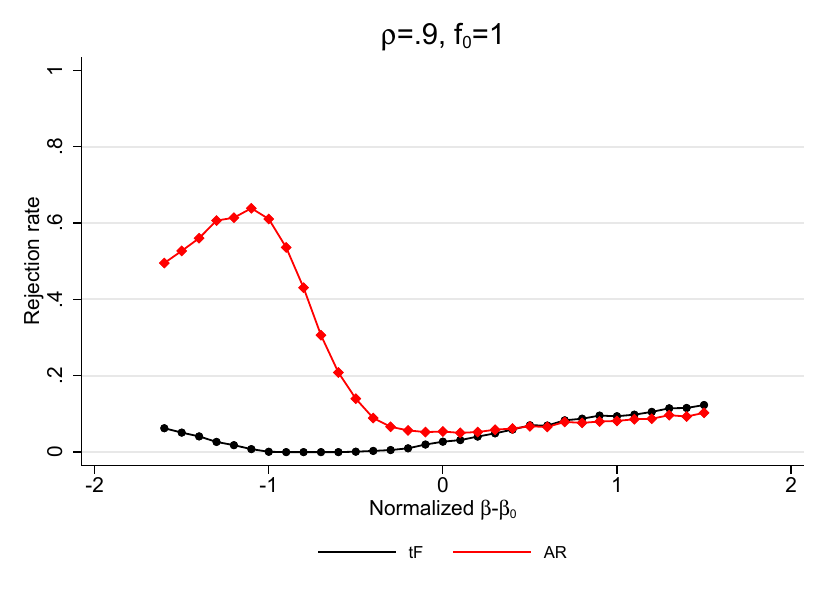}
	\end{subfigure}
\renewcommand{\thefigure}{A\arabic{figure}}
\setcounter{figure}{1}
\caption{Simulated Power Curves}
\label{fig:A1}
\end{figure}

\begin{figure}[H]
	\centering
	
	\begin{subfigure}[t]{0.375\textwidth}
		\includegraphics[width=\linewidth]{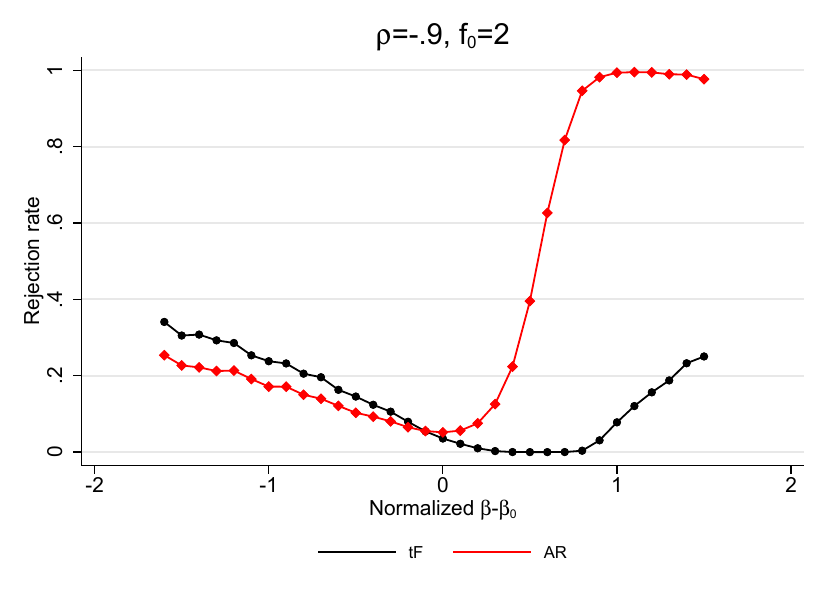}
	\end{subfigure}
	\begin{subfigure}[t]{0.375\textwidth}
		\includegraphics[width=\linewidth]{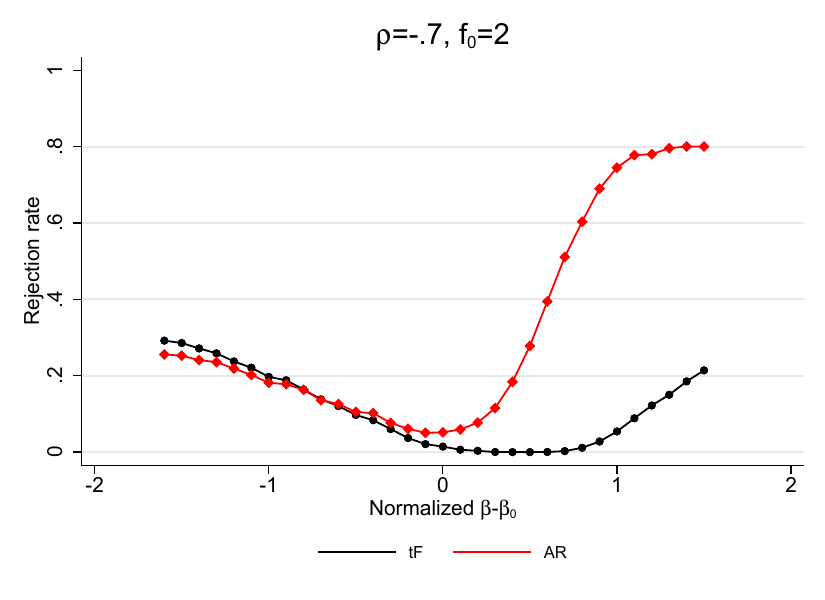}
	\end{subfigure}

	\begin{subfigure}[t]{0.375\textwidth}
		\includegraphics[width=\linewidth]{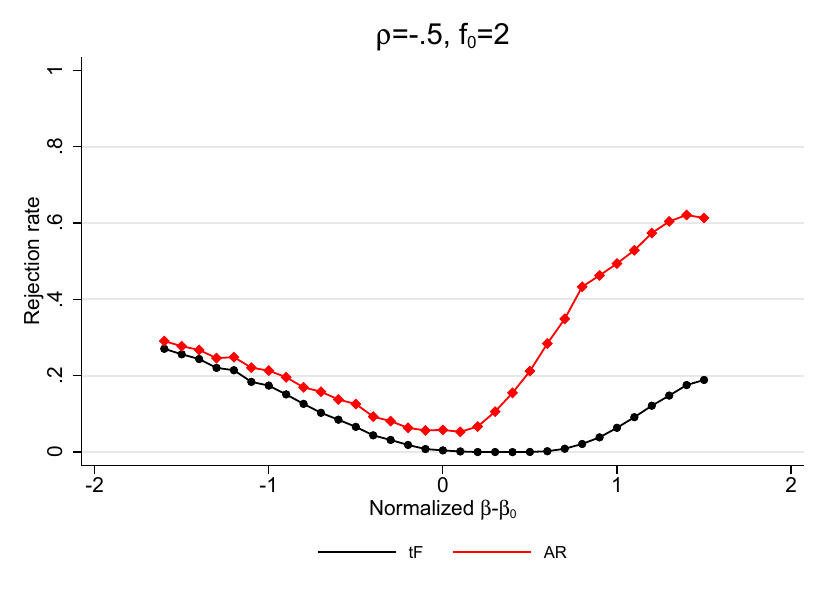}
	\end{subfigure}
	\begin{subfigure}[t]{0.375\textwidth}
		\includegraphics[width=\linewidth]{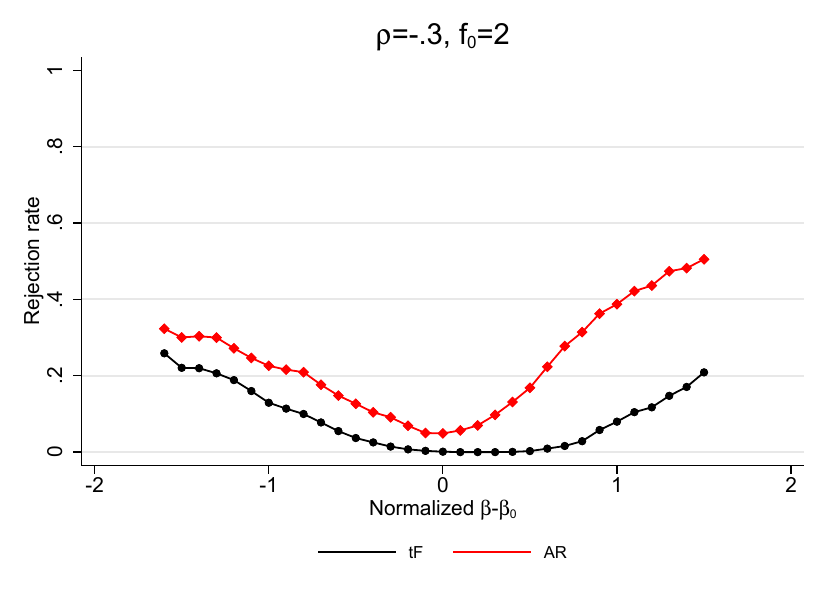}
	\end{subfigure}

	\begin{subfigure}[t]{0.375\textwidth}
		\includegraphics[width=\linewidth]{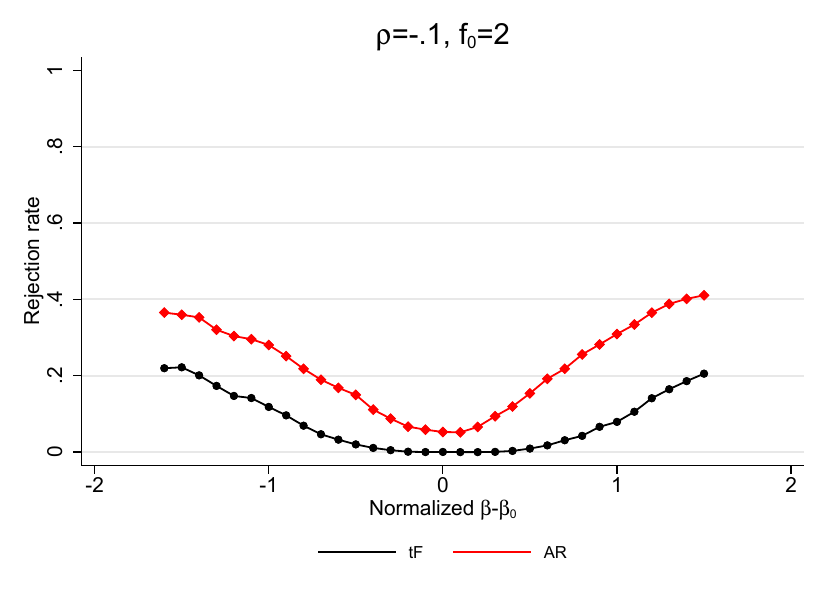}
	\end{subfigure}
    \begin{subfigure}[t]{0.375\textwidth}
		\includegraphics[width=\linewidth]{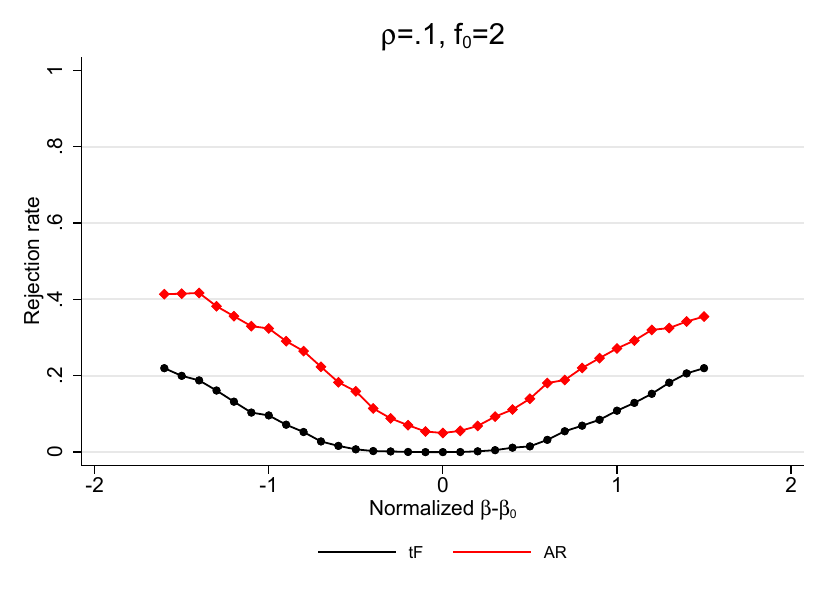}
	\end{subfigure}
    
    \begin{subfigure}[t]{0.375\textwidth}
		\includegraphics[width=\linewidth]{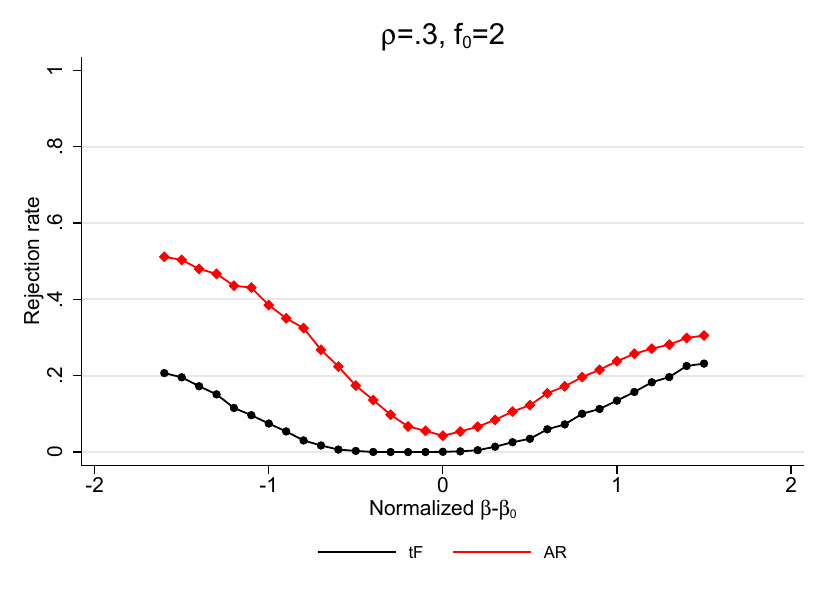}
	\end{subfigure}
    \begin{subfigure}[t]{0.375\textwidth}
		\includegraphics[width=\linewidth]{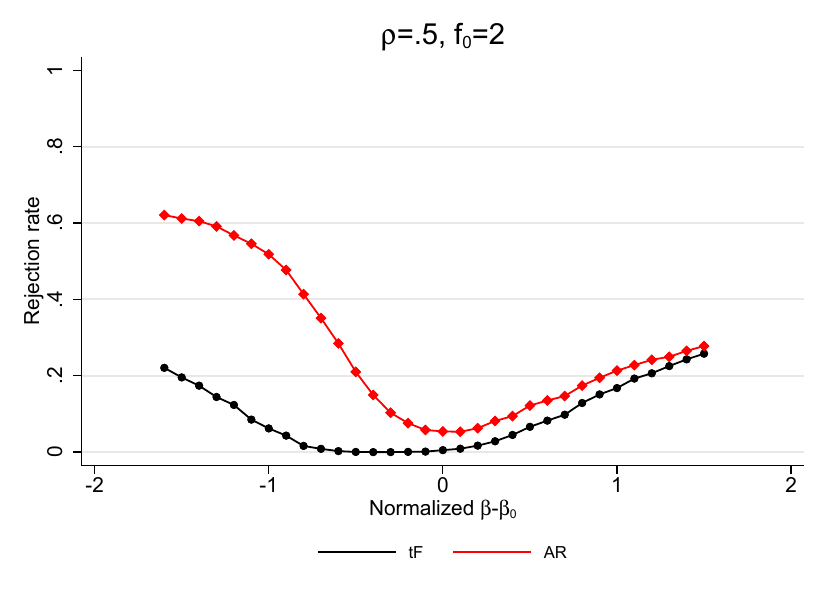}
	\end{subfigure}

    \begin{subfigure}[t]{0.375\textwidth}
		\includegraphics[width=\linewidth]{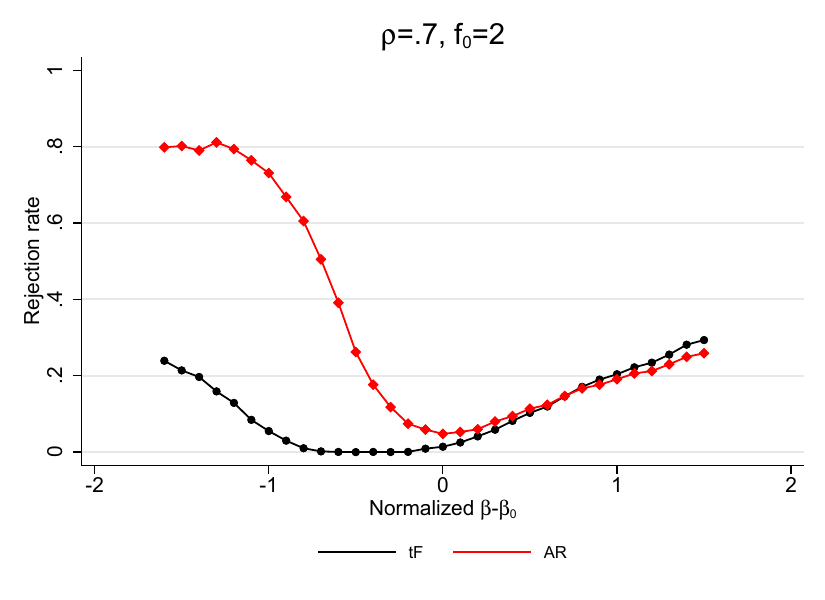}
	\end{subfigure}
    \begin{subfigure}[t]{0.375\textwidth}
		\includegraphics[width=\linewidth]{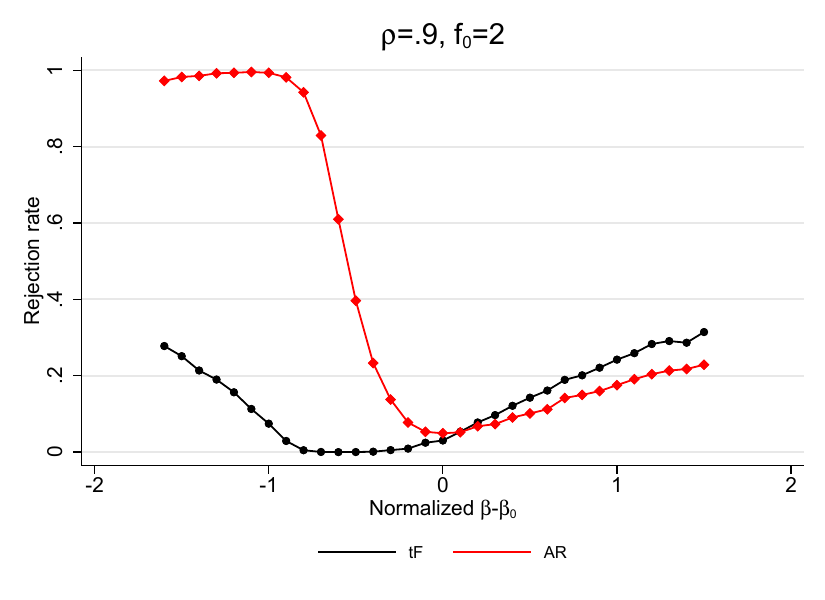}
	\end{subfigure}
\caption*{Figure A1 : Simulated Power Curves (continued)}
\end{figure}

\begin{figure}[H]
	\centering
	
	\begin{subfigure}[t]{0.375\textwidth}
		\includegraphics[width=\linewidth]{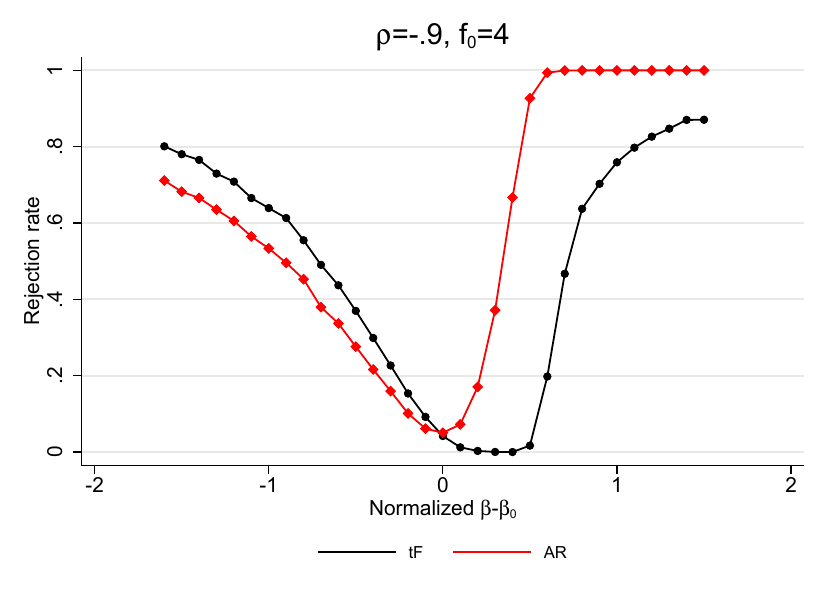}
	\end{subfigure}
	\begin{subfigure}[t]{0.375\textwidth}
		\includegraphics[width=\linewidth]{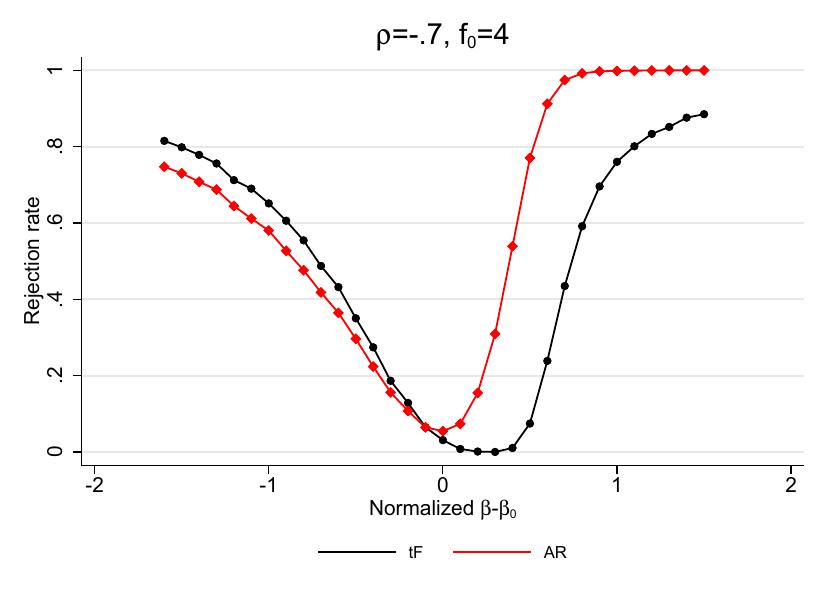}
	\end{subfigure}

	\begin{subfigure}[t]{0.375\textwidth}
		\includegraphics[width=\linewidth]{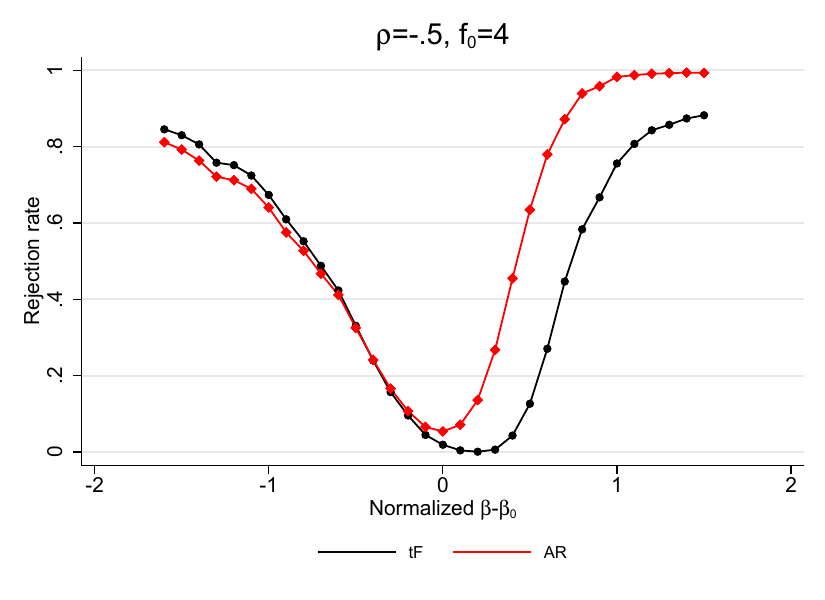}
	\end{subfigure}
	\begin{subfigure}[t]{0.375\textwidth}
		\includegraphics[width=\linewidth]{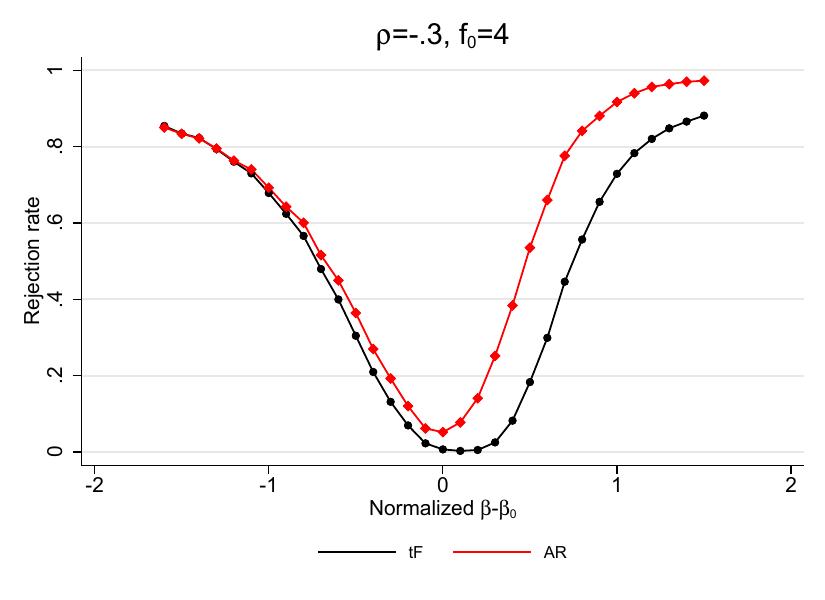}
	\end{subfigure}

	\begin{subfigure}[t]{0.375\textwidth}
		\includegraphics[width=\linewidth]{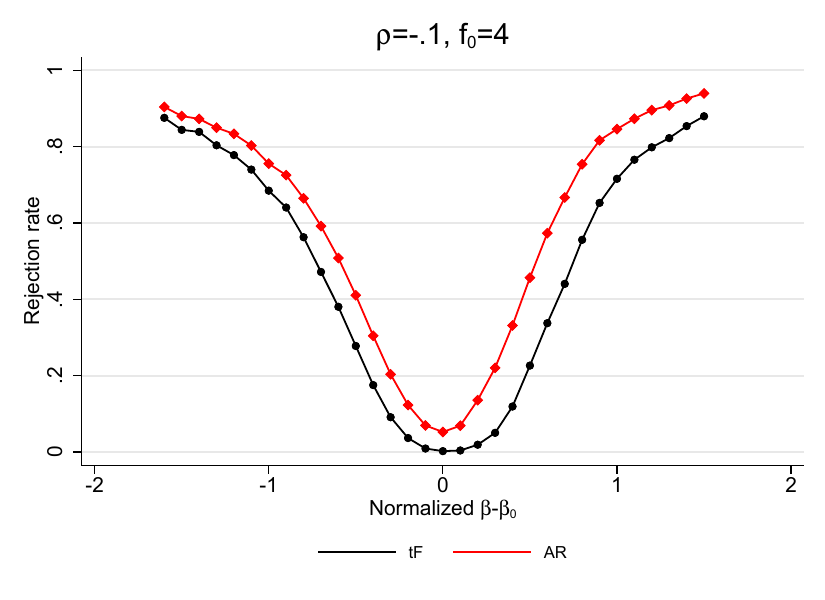}
	\end{subfigure}
    \begin{subfigure}[t]{0.375\textwidth}
		\includegraphics[width=\linewidth]{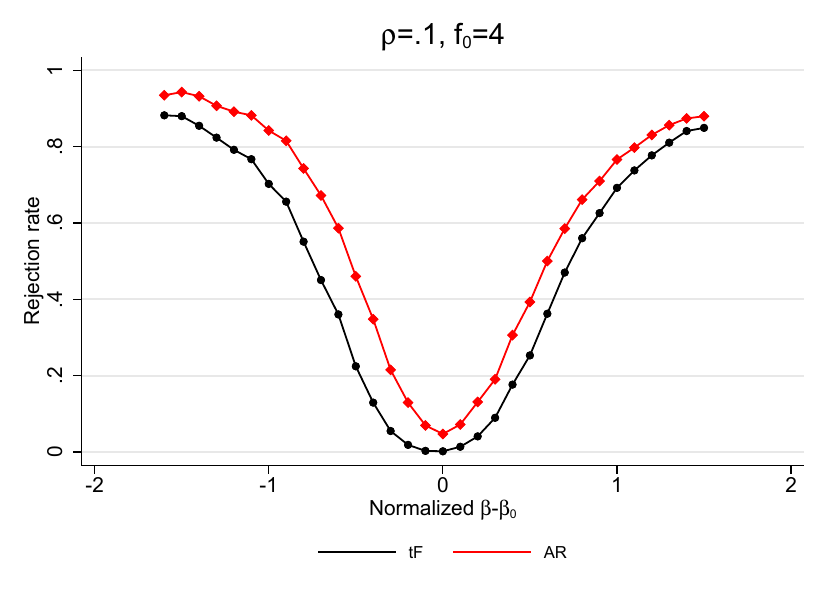}
	\end{subfigure}
    
    \begin{subfigure}[t]{0.375\textwidth}
		\includegraphics[width=\linewidth]{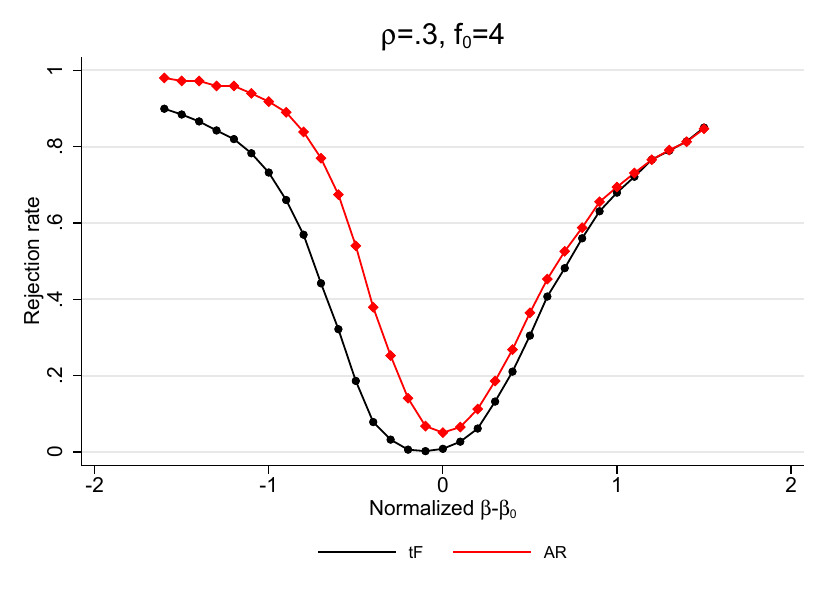}
	\end{subfigure}
    \begin{subfigure}[t]{0.375\textwidth}
		\includegraphics[width=\linewidth]{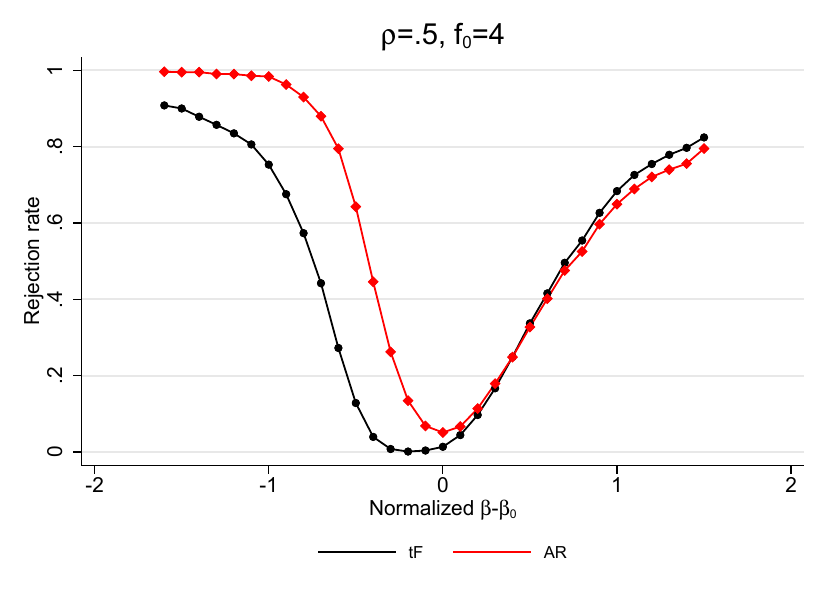}
	\end{subfigure}

    \begin{subfigure}[t]{0.375\textwidth}
		\includegraphics[width=\linewidth]{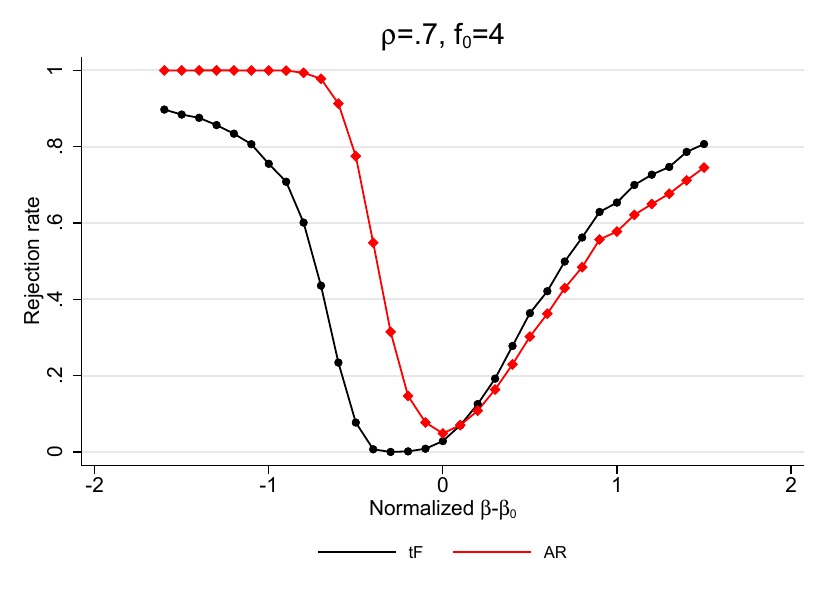}
	\end{subfigure}
    \begin{subfigure}[t]{0.375\textwidth}
		\includegraphics[width=\linewidth]{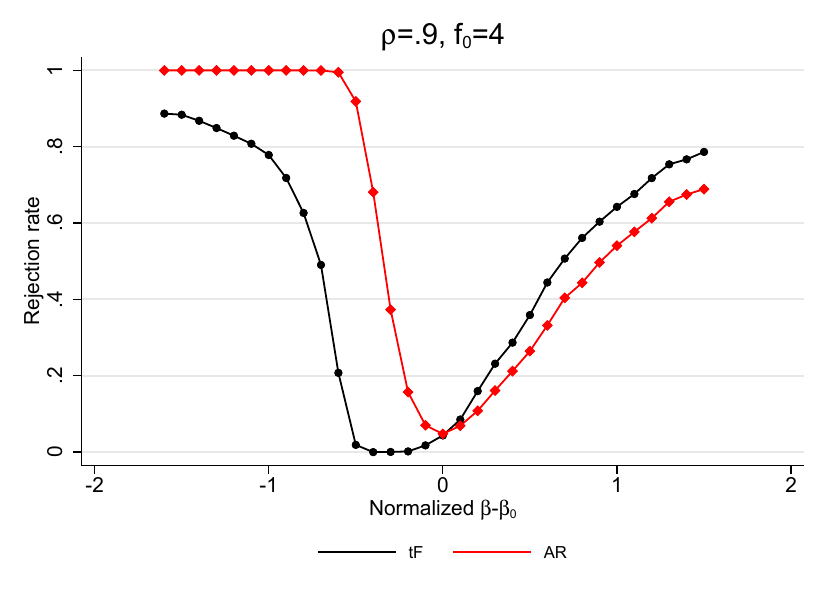}
	\end{subfigure}
\caption*{Figure A1 : Simulated Power Curves (continued)}
\end{figure}

\begin{figure}[H]
	\centering
	
	\begin{subfigure}[t]{0.375\textwidth}
		\includegraphics[width=\linewidth]{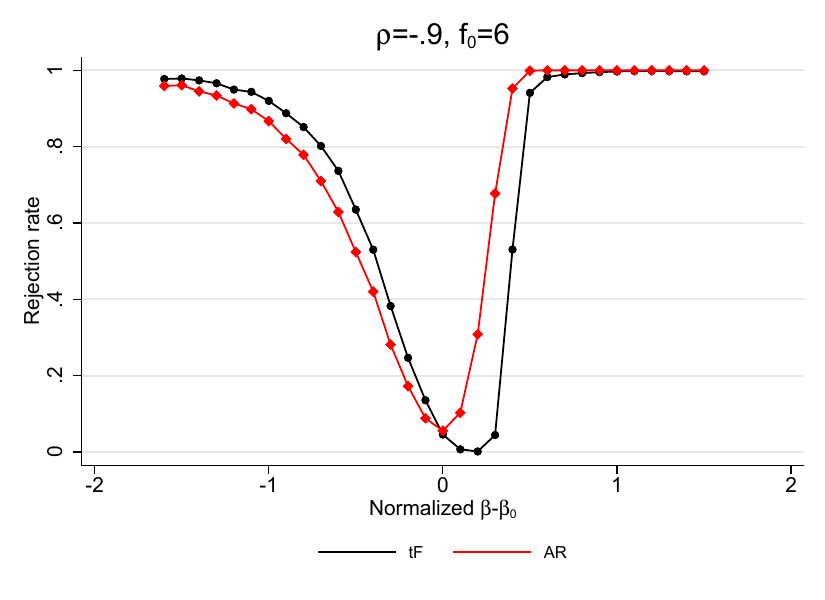}
	\end{subfigure}
	\begin{subfigure}[t]{0.375\textwidth}
		\includegraphics[width=\linewidth]{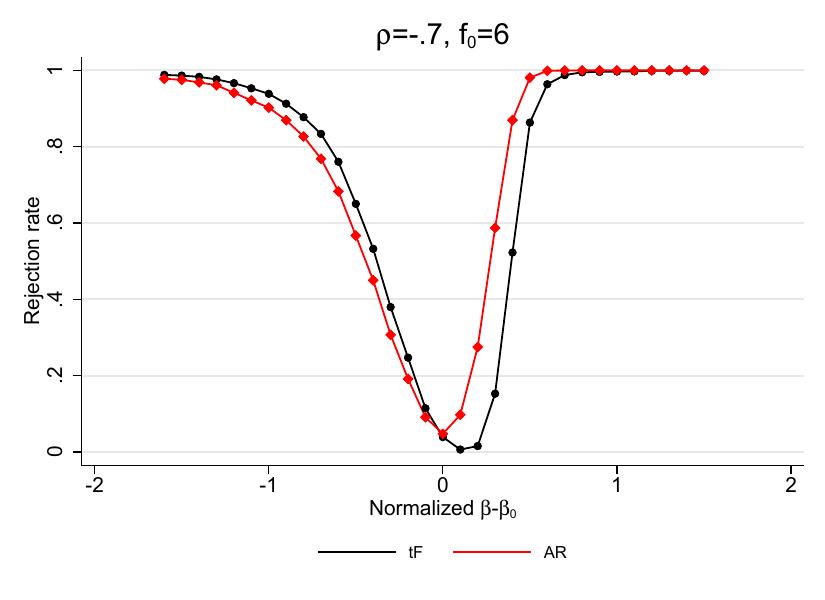}
	\end{subfigure}

	\begin{subfigure}[t]{0.375\textwidth}
		\includegraphics[width=\linewidth]{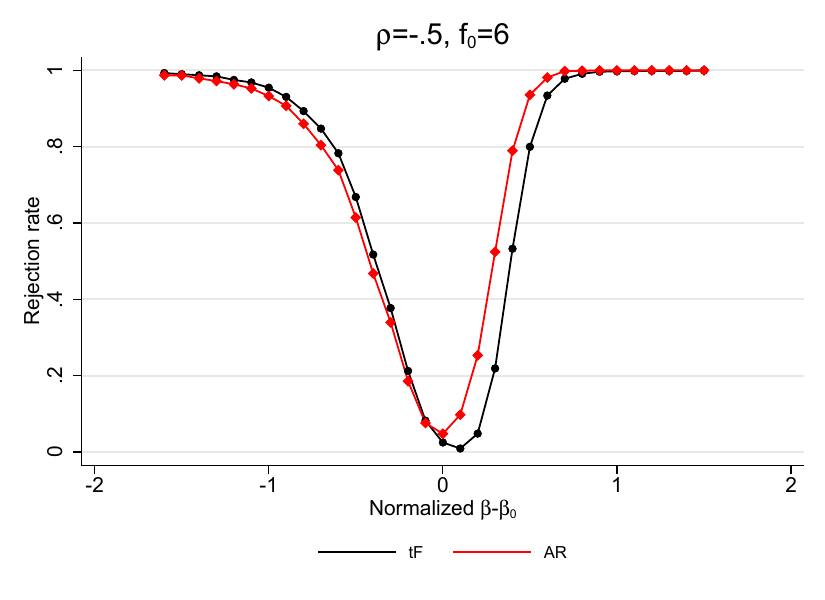}
	\end{subfigure}
	\begin{subfigure}[t]{0.375\textwidth}
		\includegraphics[width=\linewidth]{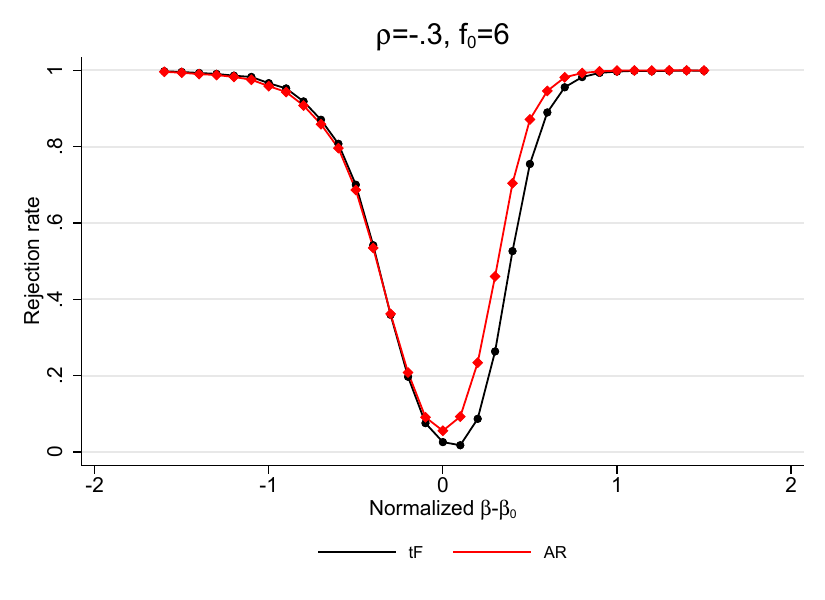}
	\end{subfigure}

	\begin{subfigure}[t]{0.375\textwidth}
		\includegraphics[width=\linewidth]{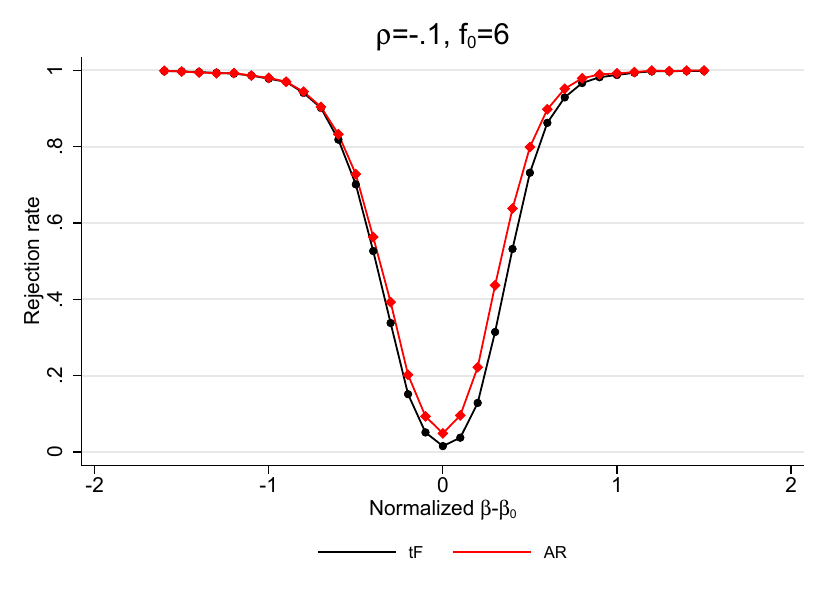}
	\end{subfigure}
    \begin{subfigure}[t]{0.375\textwidth}
		\includegraphics[width=\linewidth]{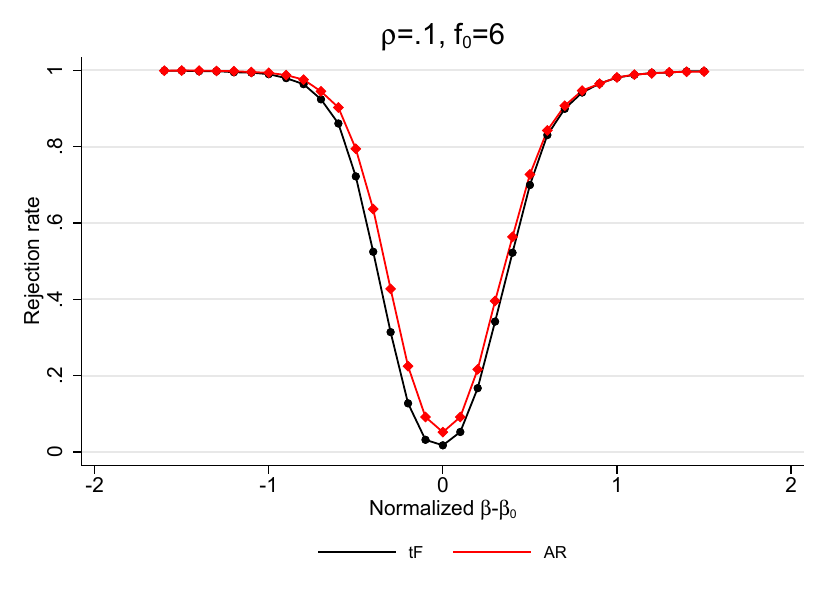}
	\end{subfigure}
    
    \begin{subfigure}[t]{0.375\textwidth}
		\includegraphics[width=\linewidth]{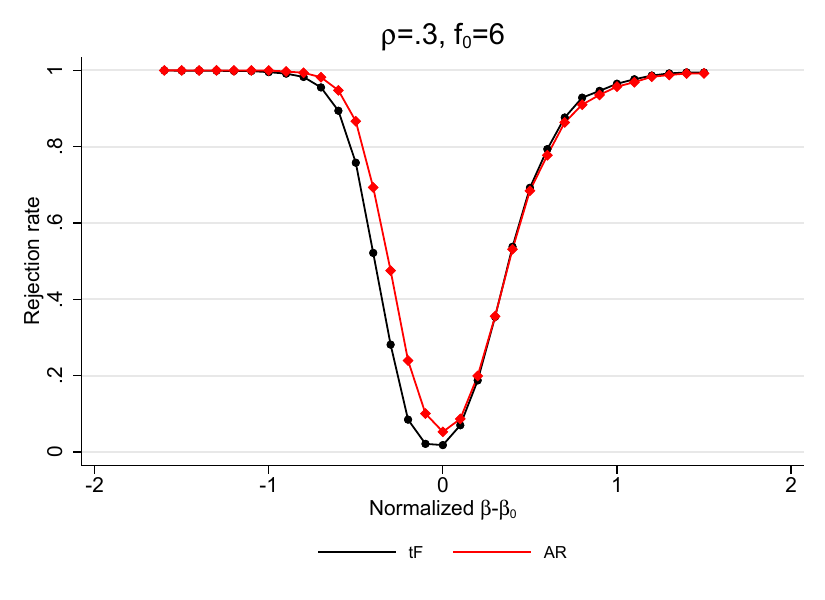}
	\end{subfigure}
    \begin{subfigure}[t]{0.375\textwidth}
		\includegraphics[width=\linewidth]{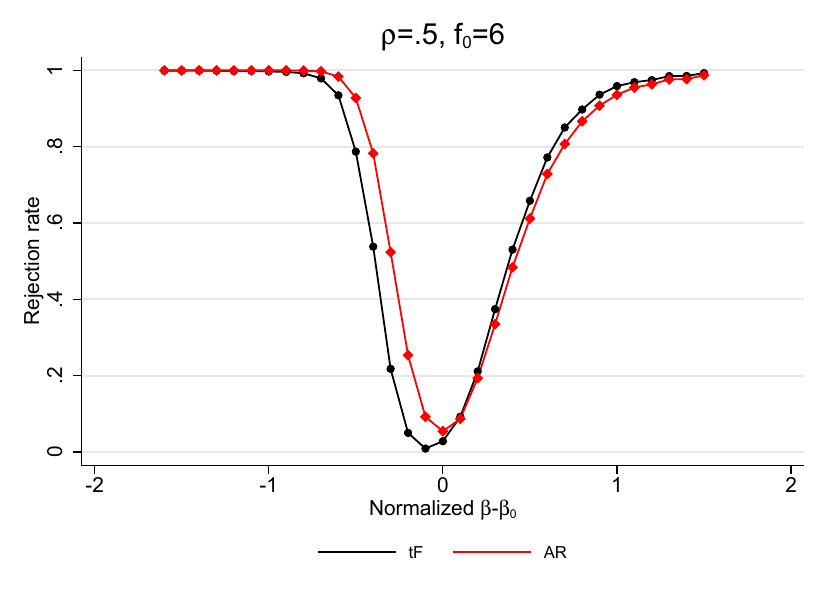}
	\end{subfigure}

    \begin{subfigure}[t]{0.375\textwidth}
		\includegraphics[width=\linewidth]{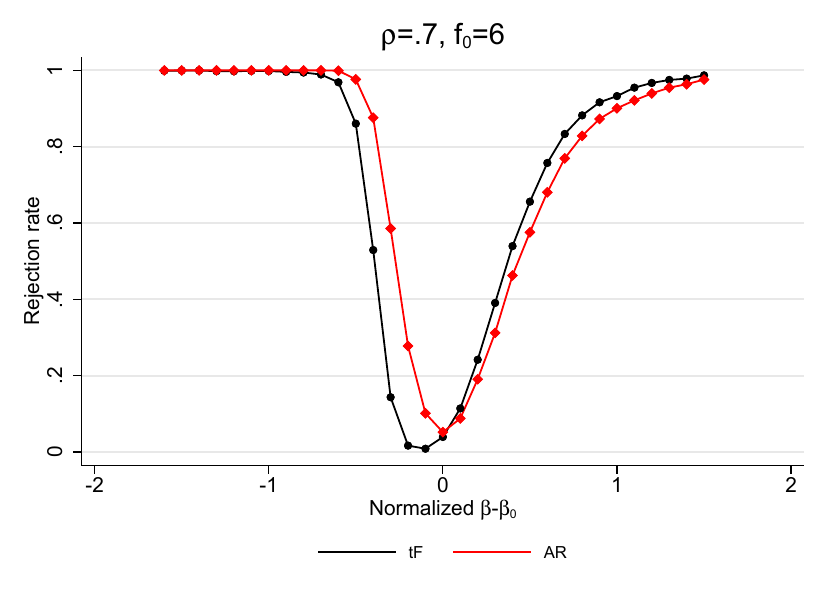}
	\end{subfigure}
    \begin{subfigure}[t]{0.375\textwidth}
		\includegraphics[width=\linewidth]{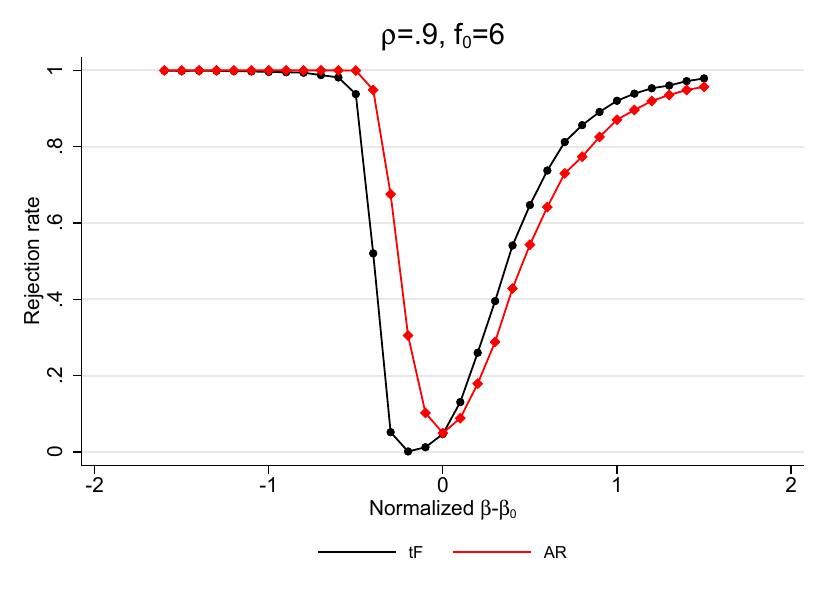}
	\end{subfigure}
\caption*{Figure A1 : Simulated Power Curves (continued)}
\end{figure}

\begin{figure}[H]
	\centering
	
	\begin{subfigure}[t]{0.375\textwidth}
		\includegraphics[width=\linewidth]{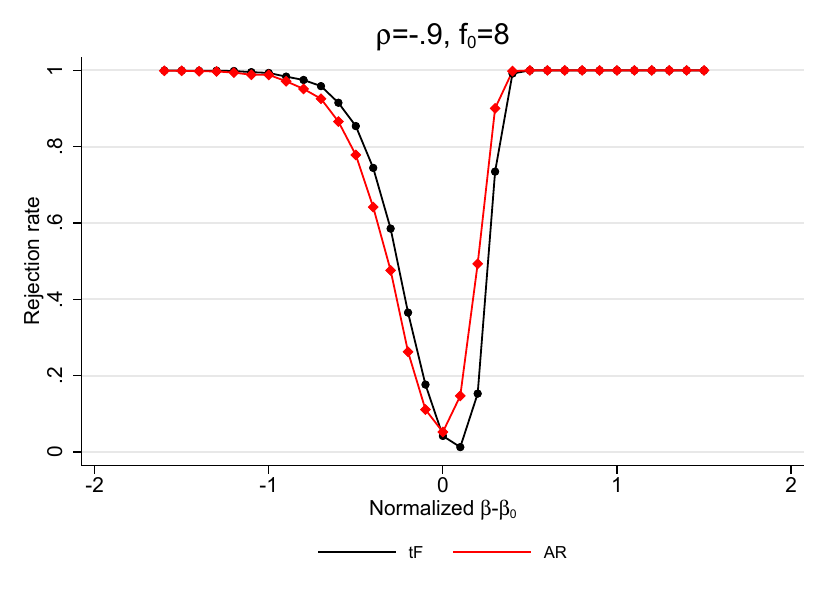}
	\end{subfigure}
	\begin{subfigure}[t]{0.375\textwidth}
		\includegraphics[width=\linewidth]{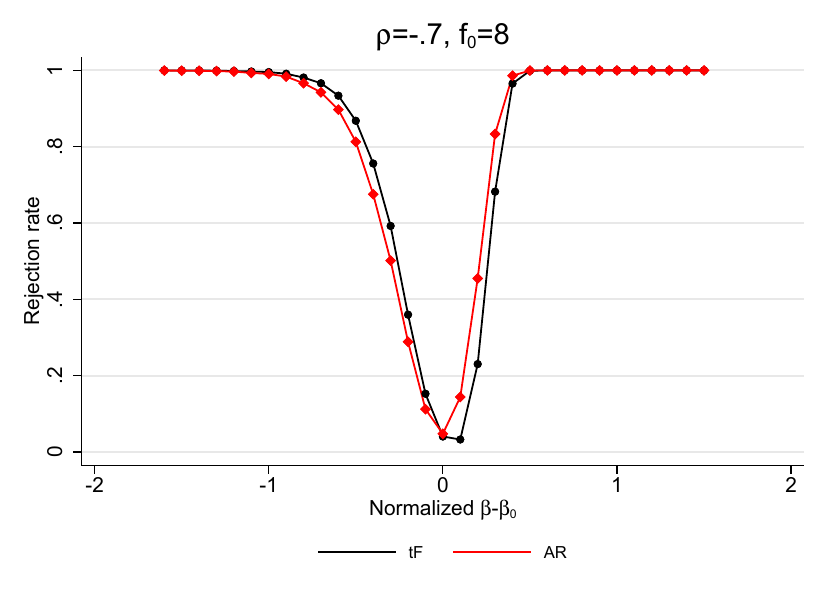}
	\end{subfigure}

	\begin{subfigure}[t]{0.375\textwidth}
		\includegraphics[width=\linewidth]{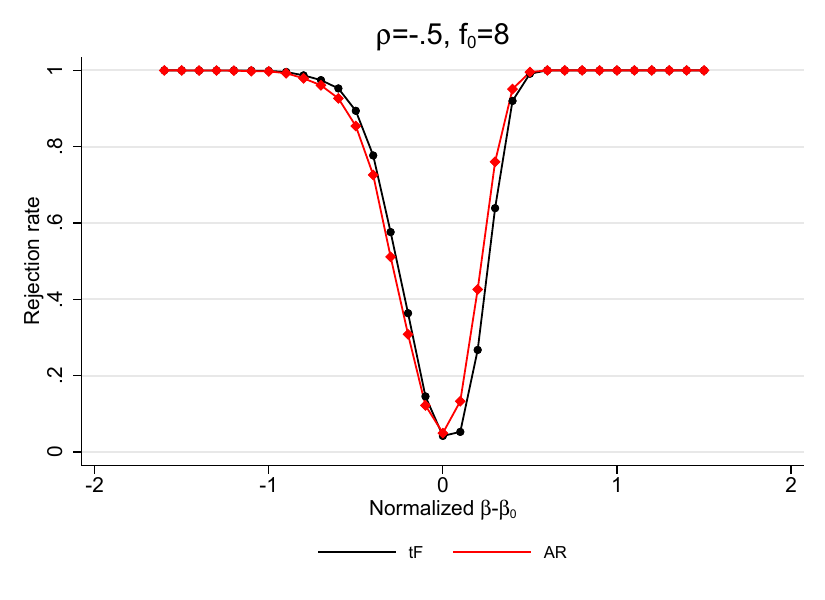}
	\end{subfigure}
	\begin{subfigure}[t]{0.375\textwidth}
		\includegraphics[width=\linewidth]{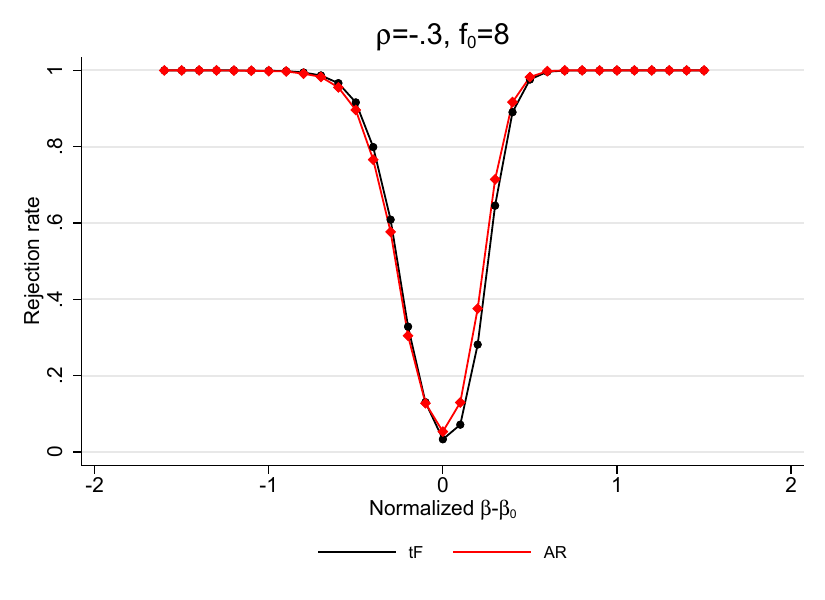}
	\end{subfigure}

	\begin{subfigure}[t]{0.375\textwidth}
		\includegraphics[width=\linewidth]{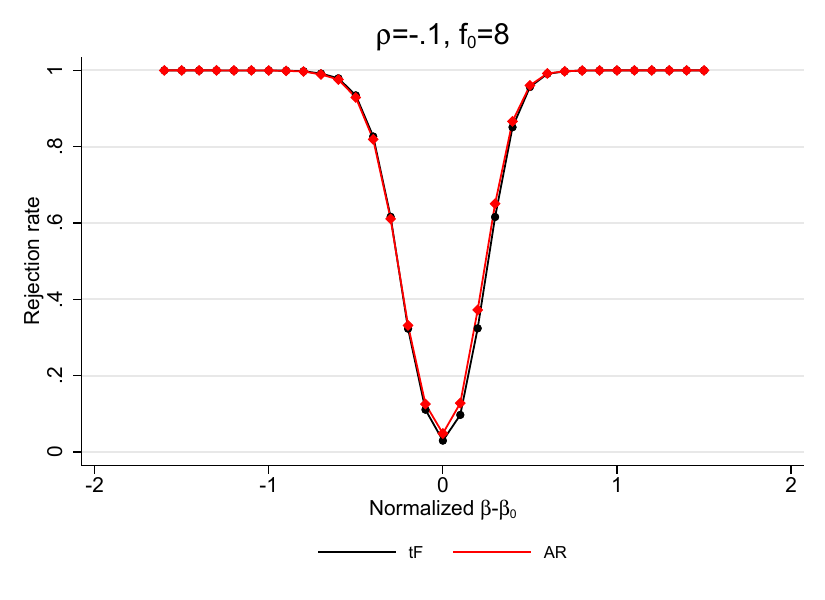}
	\end{subfigure}
    \begin{subfigure}[t]{0.375\textwidth}
		\includegraphics[width=\linewidth]{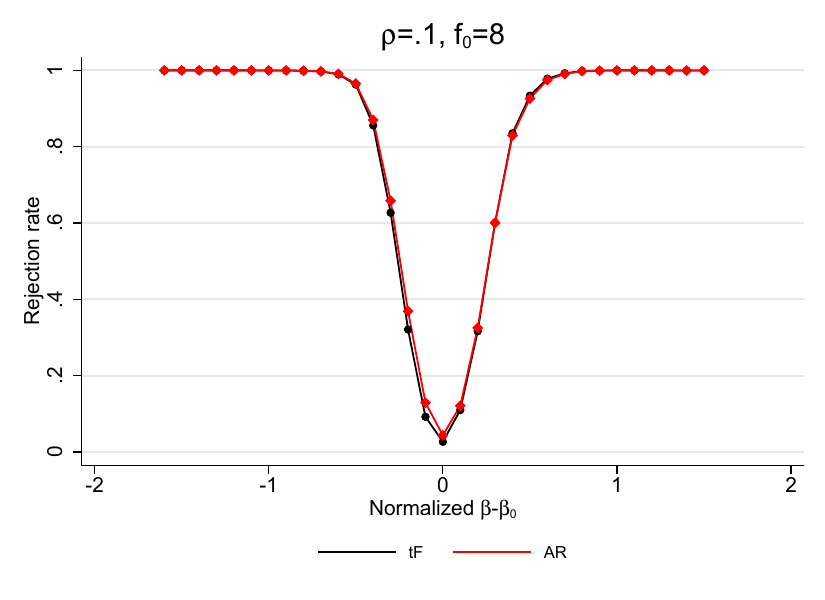}
	\end{subfigure}
    
    \begin{subfigure}[t]{0.375\textwidth}
		\includegraphics[width=\linewidth]{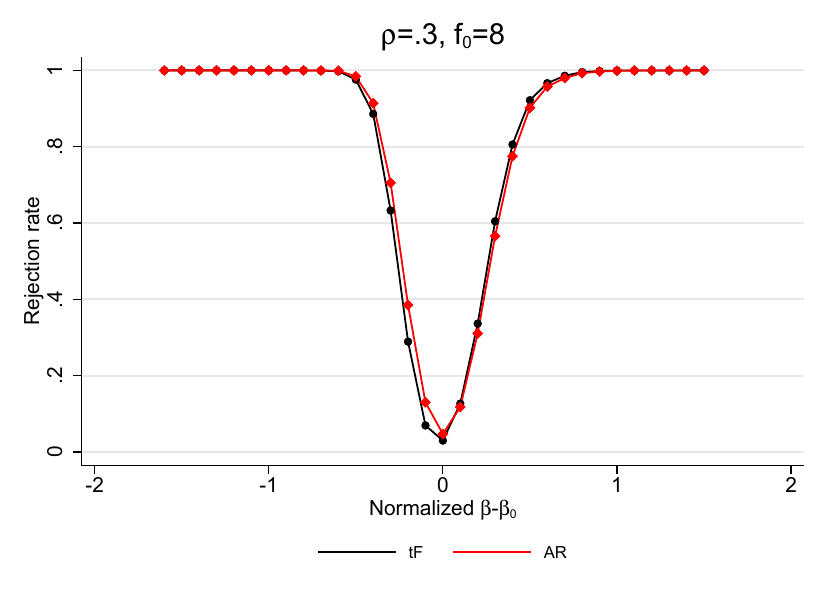}
	\end{subfigure}
    \begin{subfigure}[t]{0.375\textwidth}
		\includegraphics[width=\linewidth]{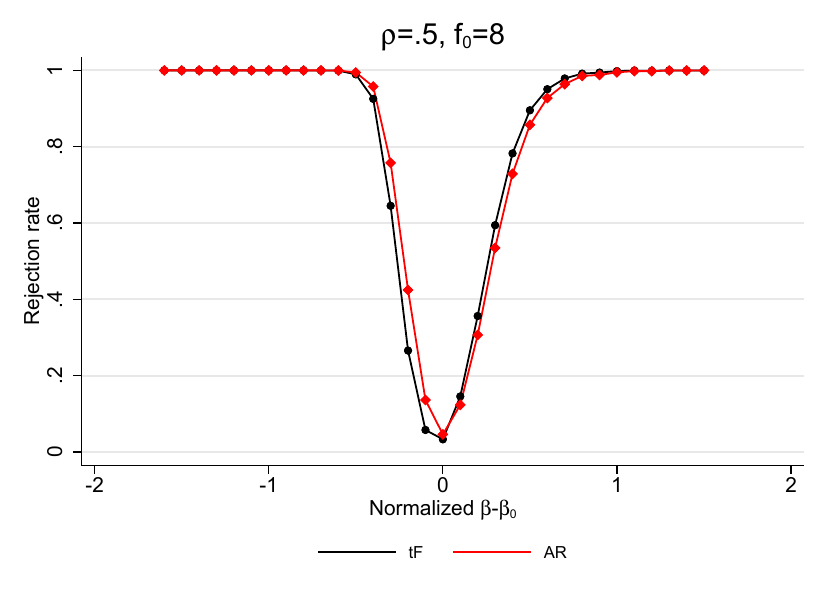}
	\end{subfigure}

    \begin{subfigure}[t]{0.375\textwidth}
		\includegraphics[width=\linewidth]{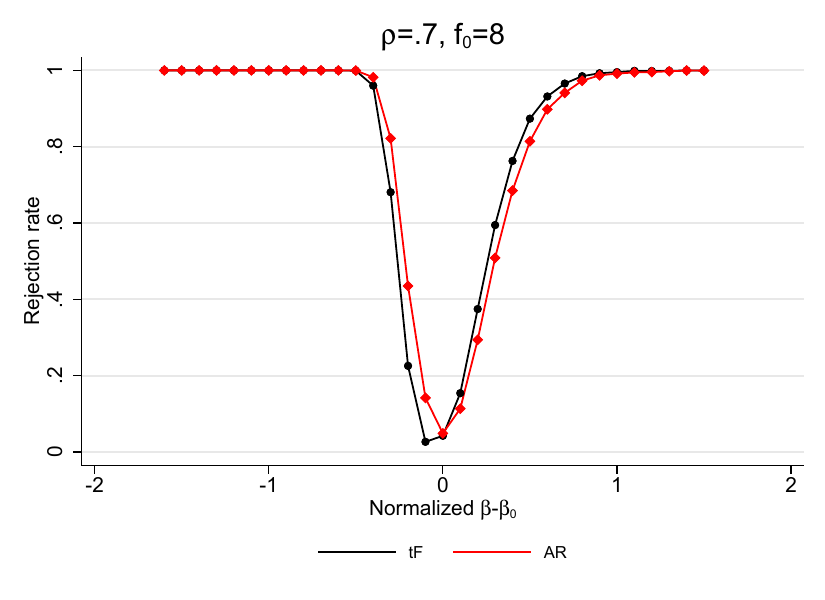}
	\end{subfigure}
    \begin{subfigure}[t]{0.375\textwidth}
		\includegraphics[width=\linewidth]{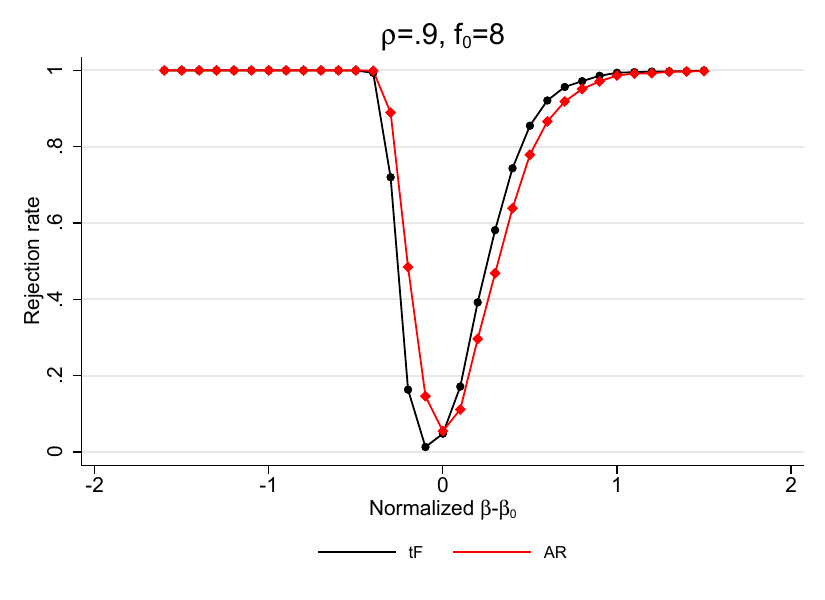}
	\end{subfigure}
\caption*{Figure A1 : Simulated Power Curves (continued)}
\end{figure}

\begin{figure}[H]
	\centering
	
	\begin{subfigure}[t]{0.375\textwidth}
		\includegraphics[width=\linewidth]{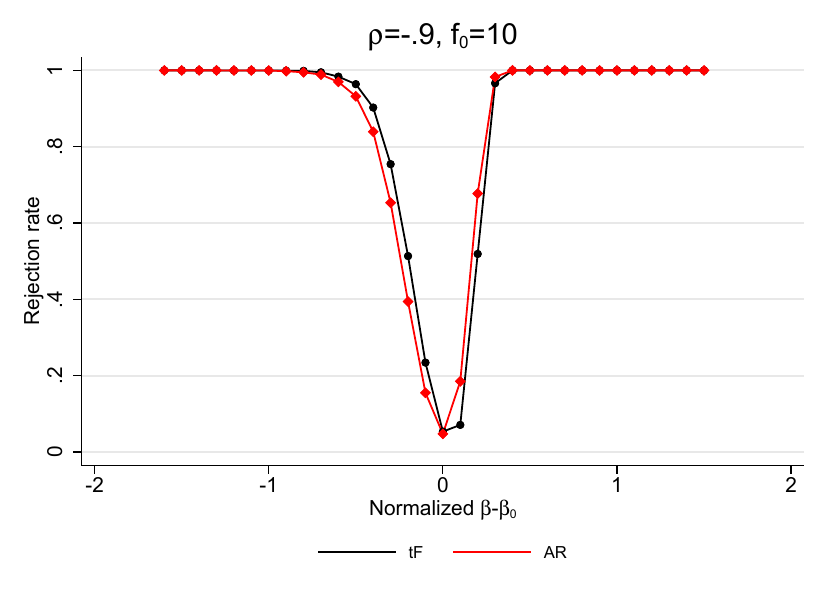}
	\end{subfigure}
	\begin{subfigure}[t]{0.375\textwidth}
		\includegraphics[width=\linewidth]{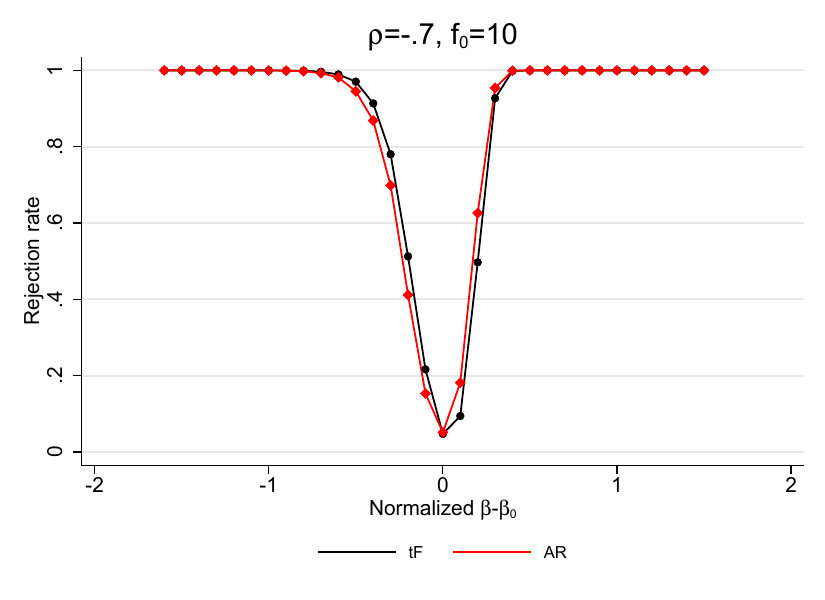}
	\end{subfigure}
    
	\begin{subfigure}[t]{0.375\textwidth}
		\includegraphics[width=\linewidth]{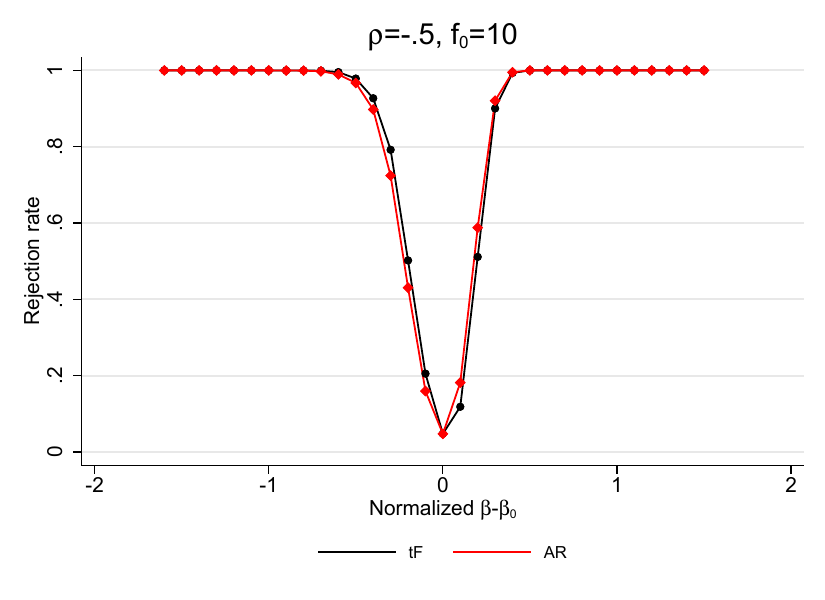}
	\end{subfigure}
	\begin{subfigure}[t]{0.375\textwidth}
		\includegraphics[width=\linewidth]{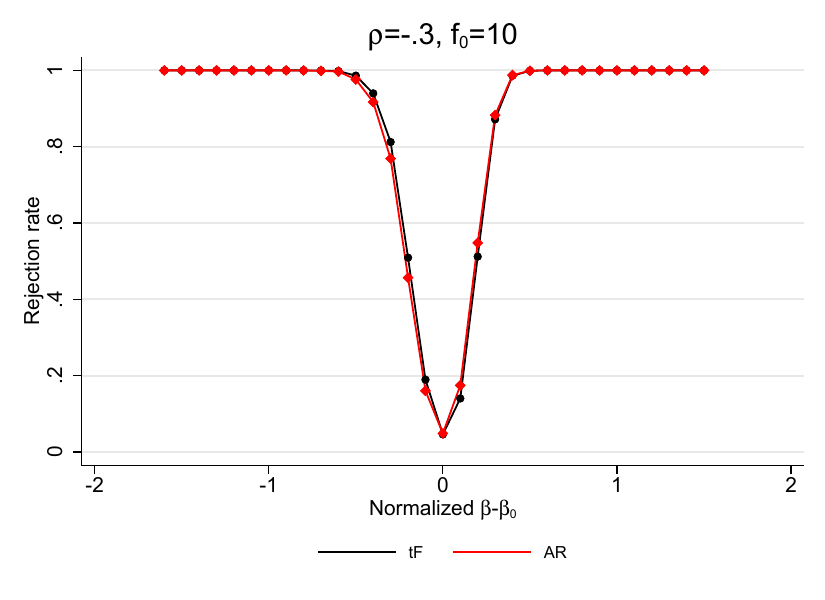}
	\end{subfigure}

	\begin{subfigure}[t]{0.375\textwidth}
		\includegraphics[width=\linewidth]{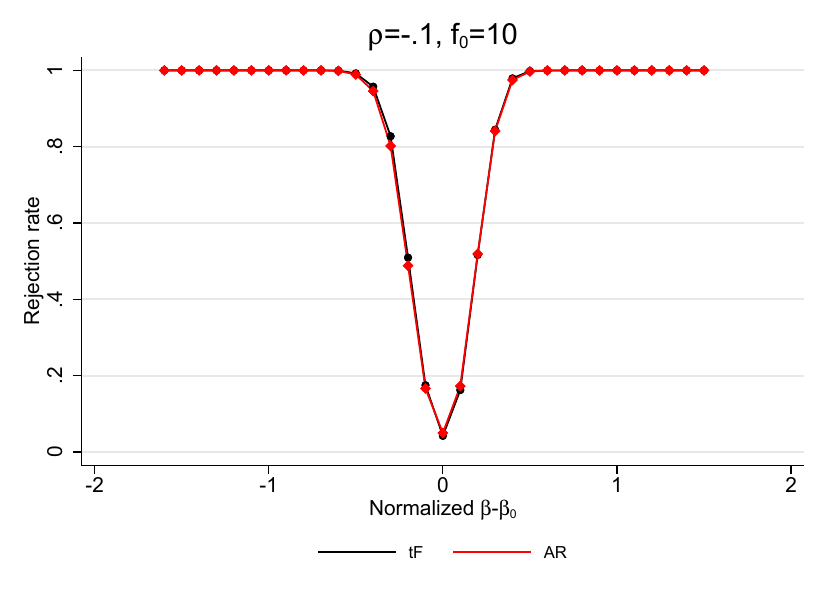}
	\end{subfigure}
    \begin{subfigure}[t]{0.375\textwidth}
		\includegraphics[width=\linewidth]{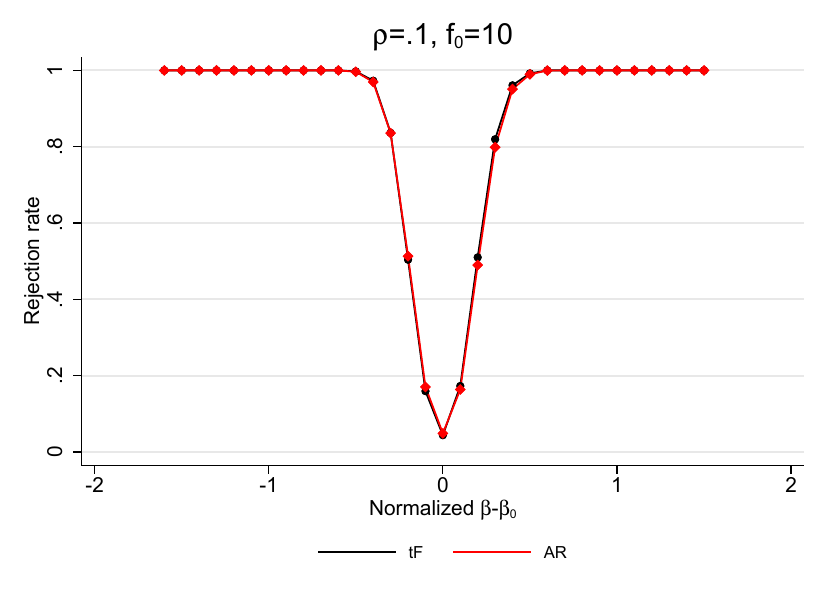}
	\end{subfigure}
    
    \begin{subfigure}[t]{0.375\textwidth}
		\includegraphics[width=\linewidth]{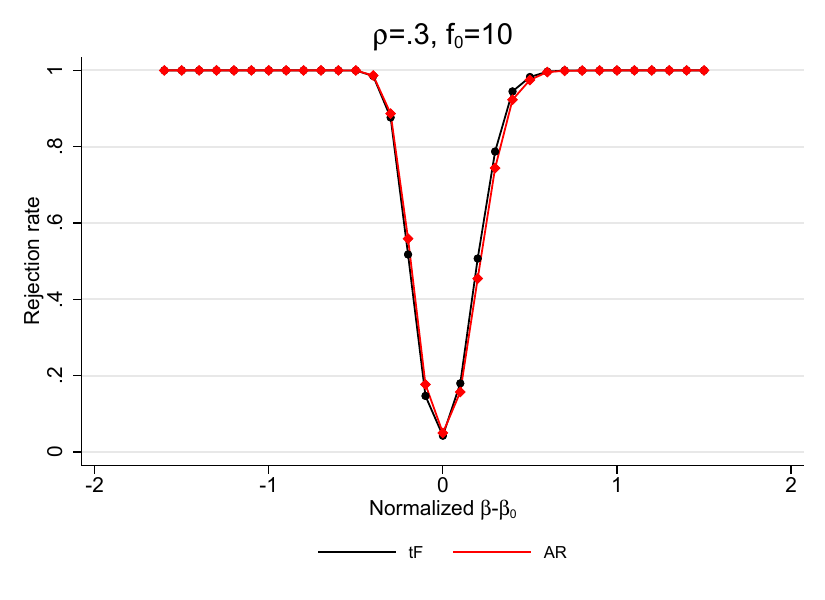}
	\end{subfigure}
    \begin{subfigure}[t]{0.375\textwidth}
		\includegraphics[width=\linewidth]{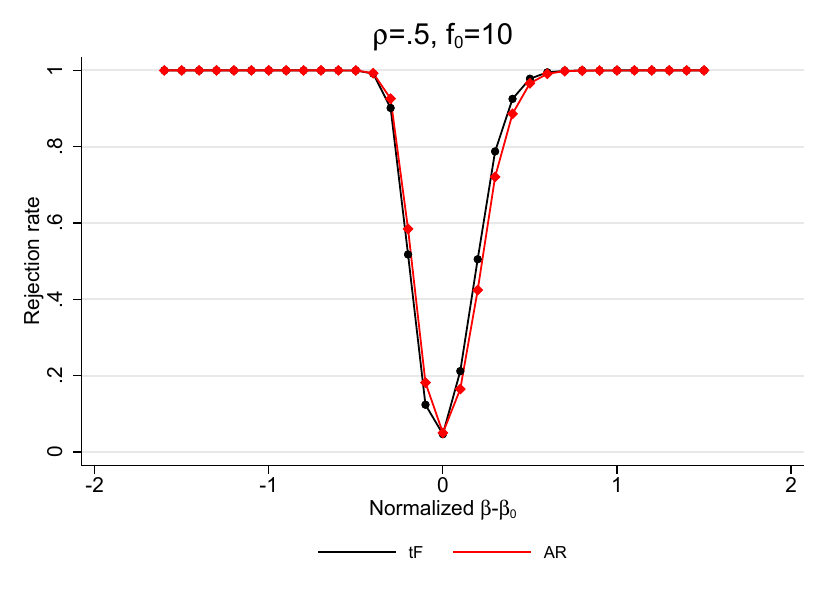}
	\end{subfigure}

    \begin{subfigure}[t]{0.375\textwidth}
		\includegraphics[width=\linewidth]{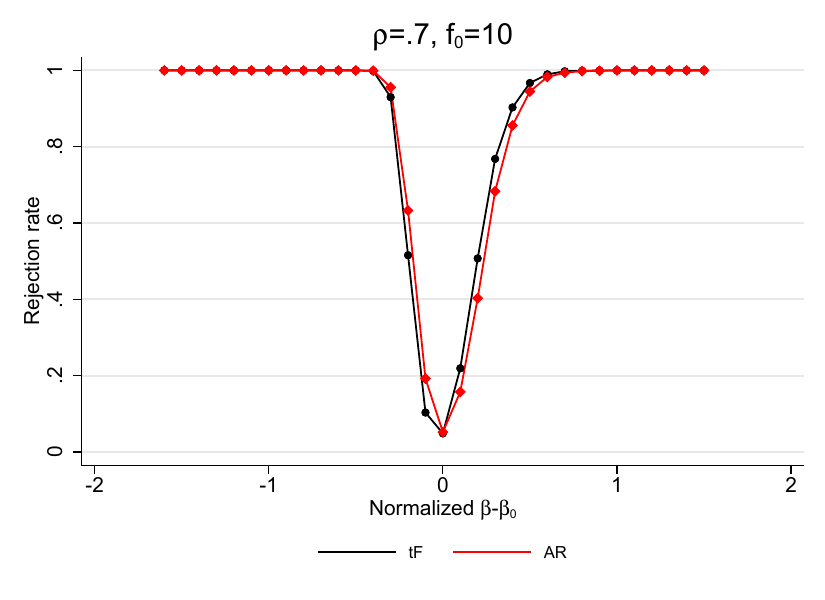}
	\end{subfigure}
    \begin{subfigure}[t]{0.375\textwidth}
		\includegraphics[width=\linewidth]{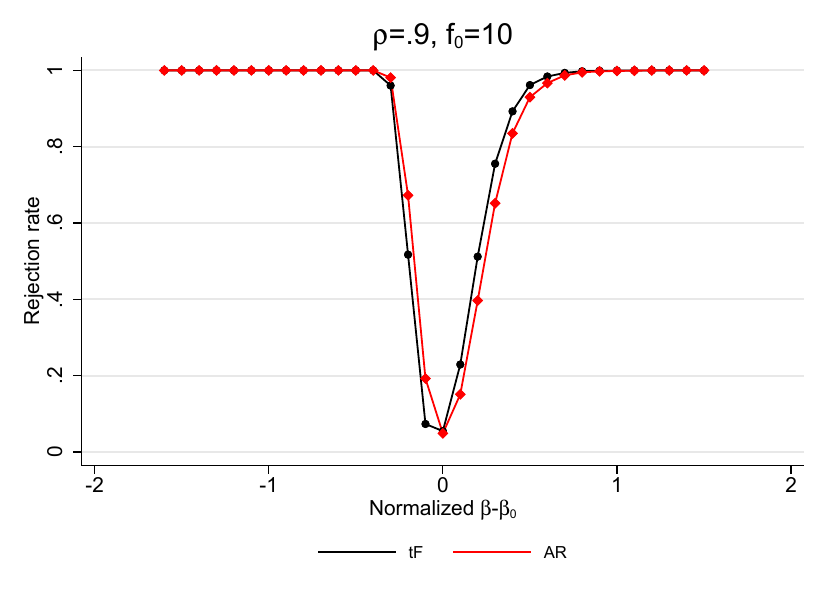}
	\end{subfigure}
\caption*{Figure A1 : Simulated Power Curves (continued)}
\end{figure}

\subsection*{B \quad AER Paper Details}
\pdfbookmark[2]{AER Paper Details}{AER}

\begin{enumerate}[label=\textbf{\alph*.}]
    \item \citet{rosenthal2014private}

    Number of Specifications: 2
    
    JEL Classification:\\
    R21 Urban, Rural, Regional, Real Estate, and Transportation Economics: Housing Demand\\
    R31 Housing Supply and Markets\\
    R38 Production Analysis and Firm Location: Government Policy
    
    \item \citet{di2015conveniently}

    Number of Specifications: 4
    
    JEL Classification:\\
    C72 Noncooperative Games\\
    D63 Equity, Justice, Inequality, and Other Normative Criteria and Measurement\\
    D64 Altruism; Philanthropy\\
    D83 Search; Learning; Information and Knowledge; Communication; Belief; Unawareness
    
    \item \citet{field2013does}

    Number of Specifications: 4

    JEL Classification:\\
    A21 Economic Education and Teaching of Economics: Pre-college\\
    G32 Financing Policy; Financial Risk and Risk Management;Capital and Ownership Structure; Value of Firms; Goodwill\\
    I32 Measurement and Analysis of Poverty\\
    L25 Firm Performance: Size, Diversification, and Scope\\
    L26 Entrepreneurship\\
    O15 Economic Development: Human Resources; Human Development; Income Distribution; Migration\\
    O16 Economic Development: Financial Markets; Saving and Capital Investment; Corporate Finance and Governance

    \item \citet{rao2019familiarity}

    Number of Specifications: 12
    
    JEL Classification:\\
    C90 Design of Experiments: General\\
    D31 Personal Income, Wealth, and Their Distributions\\
    I21 Analysis of Education\\
    I24 Education and Inequality\\
    O15 Economic Development: Human Resources; Human Development; Income Distribution; Migration\\
    Z13 Economic Sociology; Economic Anthropology; Language; Social and Economic Stratification

    \item \citet{moser2014german}

    Number of Specifications: 10

    JEL Classification:\\
    J15 Economics of Minorities, Races, Indigenous Peoples, and Immigrants; Non-labor Discrimination\\
    L65 Chemicals; Rubber; Drugs; Biotechnology\\
    N62 Economic History: Manufacturing and Construction: U.S.; Canada: 1913-\\
    O31 Innovation and Invention: Processes and Incentives\\
    O34 Intellectual Property and Intellectual Capital

    \item \citet{voigtlander2013west}

    Number of Specifications: 5
    
    JEL Classification:\\
    J12 Marriage; Marital Dissolution; Family Structure; Domestic Abuse\\
    J13 Fertility; Family Planning; Child Care; Children; Youth\\
    J16 Economics of Gender; Non-labor Discrimination\\
    N33 Economic History: Labor and Consumers, Demography, Education, Health, Welfare, Income, Wealth, Religion, and Philanthropy: Europe: Pre-1913\\
    N53 Economic History: Agriculture, Natural Resources, Environment, and Extractive Industries: Europe: Pre-1913\\
    Q11 Agriculture: Aggregate Supply and Demand Analysis; Prices

    \item \citet{hornung2014immigration}

    Number of Specifications: 4

    JEL Classification:\\
    J24 Human Capital; Skills; Occupational Choice; Labor Productivity\\
    J61 Geographic Labor Mobility; Immigrant Workers\\
    L67 Other Consumer Nondurables\\
    N33 Economic History: Labor and Consumers, Demography, Education, Health, Welfare, Income, Wealth, Religion, and Philanthropy: Europe: Pre-1913\\
    N63 Economic History: Manufacturing and Construction: Europe: Pre-1913\\
    O33 Technological Change: Choices and Consequences; Diffusion Processes\\
    O47 Measurement of Economic Growth; Aggregate Productivity; Cross-Country Output Convergence

    \item \citet{ottaviano2013immigration}

    Number of Specifications: 12

    JEL Classification:\\
    J24 Human Capital; Skills; Occupational Choice; Labor Productivity\\
    J41 Labor Contracts\\
    J61 Geographic Labor Mobility; Immigrant Workers\\
    L24 Contracting Out; Joint Ventures; Technology Licensing

    \item \citet{alesina2018intergenerational}

    Number of Specifications: 5
    
    JEL Classification:\\
    D63 Equity, Justice, Inequality, and Other Normative Criteria and Measurement\\
    D72 Political Processes: Rent-seeking, Lobbying, Elections, Legislatures, and Voting Behavior\\
    H23 Taxation and Subsidies: Externalities; Redistributive Effects; Environmental Taxes and Subsidies\\
    H24 Personal Income and Other Nonbusiness Taxes and Subsidies; includes inheritance and gift taxes\\
    J31 Wage Level and Structure; Wage Differentials\\
    J62 Job, Occupational, and Intergenerational Mobility; Promotion

    \item \citet{campante2014isolated}

    Number of Specifications: 4
    
    JEL Classification:\\
    D72 Political Processes: Rent-seeking, Lobbying, Elections, Legislatures, and Voting Behavior\\
    D73 Bureaucracy; Administrative Processes in Public Organizations; Corruption\\
    H41 Public Goods\\
    H83 Public Administration; Public Sector Accounting and Audits\\
    K42 Illegal Behavior and the Enforcement of Law\\
    R23 Urban, Rural, Regional, Real Estate, and Transportation Economics: Regional Migration; Regional Labor Markets; Population; Neighborhood Characteristics

    \item \citet{decarolis2015medicare}
    
    Number of Specifications: 6
    
    JEL Classification:\\
    G22 Insurance; Insurance Companies; Actuarial Studies\\
    H51 National Government Expenditures and Health\\
    I13 Health Insurance, Public and Private\\
    I18 Health: Government Policy; Regulation; Public Health

    \item \citet{fabra2014pass}

    Number of Specifications: 5

    JEL Classification:\\
    D44 Auctions\\
    L11 Production, Pricing, and Market Structure; Size Distribution of Firms\\
    L94 Electric Utilities\\
    L98 Industry Studies: Utilities and Transportation: Government Policy\\
    Q52 Pollution Control Adoption Costs; Distributional Effects; Employment Effects\\
    Q54 Climate; Natural Disasters; Global Warming

    \item \citet{steinwender2018real}

    Number of Specifications: 8

    JEL Classification:\\
    D83 Search; Learning; Information and Knowledge; Communication; Belief; Unawareness\\
    F12 Models of Trade with Imperfect Competition and Scale Economies; Fragmentation\\
    F14 Empirical Studies of Trade\\
    L96 Telecommunications\\
    N71 Economic History: Transport, Trade, Energy, Technology, and Other Services: U.S.; Canada: Pre-1913\\
    N73 Economic History: Transport, Trade, Energy, Technology, and Other Services: Europe: Pre-1913

    \item \citet{condra2018logic}
    
    Number of Specifications: 9

    JEL Classification:\\
    D72 Political Processes: Rent-seeking, Lobbying, Elections, Legislatures, and Voting Behavior\\
    D74 Conflict; Conflict Resolution; Alliances; Revolutions\\
    O17 Formal and Informal Sectors; Shadow Economy; Institutional Arrangements

    \item \citet{brollo2013political}

    Number of Specifications: 10

    JEL Classification:\\
    D72 Political Processes: Rent-seeking, Lobbying, Elections, Legislatures, and Voting Behavior\\
    D73 Bureaucracy; Administrative Processes in Public Organizations; Corruption\\
    H77 Intergovernmental Relations; Federalism; Secession\\
    O17 Formal and Informal Sectors; Shadow Economy; Institutional Arrangements\\
    O18 Economic Development: Urban, Rural, Regional, and Transportation Analysis; Housing; Infrastructure

    \item \citet{berman2017mine}

    Number of Specifications: 3
    
    JEL Classification:\\
    C23 Single Equation Models; Single Variables: Panel Data Models; Spatio-temporal Models\\
    D74 Conflict; Conflict Resolution; Alliances; Revolutions\\
    L70 Industry Studies: Primary Products and Construction: General\\
    O13 Economic Development: Agriculture; Natural Resources; Energy; Environment; Other Primary Products\\
    Q34 Natural Resources and Domestic and International Conflicts

    \item \citet{nunn2014us}

    Number of Specifications: 48

    JEL Classification:\\
    D74 Conflict; Conflict Resolution; Alliances\\
    F35 Foreign Aid\\
    O17 Formal and Informal Sectors; Shadow Economy; Institutional Arrangements\\
    O19 International Linkages to Development; Role of International Organizations\\
    Q11 Agriculture: Aggregate Supply and Demand Analysis; Prices\\
    Q18 Agricultural Policy; Food Policy

\end{enumerate}

\end{document}